\documentclass{article} %
\usepackage{iclr2026_conference,times}

\usepackage{amsmath,amsfonts,bm}

\def\eqref#1{equation~\ref{#1}}

\def\1{\bm{1}}

\def\va{{\bm{a}}}

\def\vm{{\bm{m}}}

\def\vs{{\bm{s}}}

\def\vx{{\bm{x}}}

\def\vz{{\bm{z}}}

\def\mW{{\bm{W}}}

\DeclareMathAlphabet{\mathsfit}{\encodingdefault}{\sfdefault}{m}{sl}
\SetMathAlphabet{\mathsfit}{bold}{\encodingdefault}{\sfdefault}{bx}{n}

\usepackage{pgfplots}
\pgfplotsset{compat=1.18}

\usepackage{epsfig}
\usepackage{stfloats}
\usepackage{subcaption}
\usepackage{makecell}
\usepackage{lipsum}  

\usepackage[utf8]{inputenc} %
\usepackage[T1]{fontenc}    %
\definecolor{citecolor}{HTML}{0071bc}
\usepackage[pagebackref,breaklinks=true,colorlinks,citecolor=citecolor,bookmarks=false]{hyperref}
\usepackage{url}            %
\usepackage{booktabs}       %
\usepackage{amsfonts}       %
\usepackage{nicefrac}       %
\usepackage{microtype}      %
\usepackage[dvipsnames,table]{xcolor}
\usepackage{graphicx}
\usepackage{amsmath}
\usepackage{algorithm}
\usepackage{algorithmic}
\usepackage{comment}
\usepackage{bbm}
\usepackage{dsfont}
\usepackage{easylist}
\usepackage{xspace}
\usepackage{wrapfig}
\usepackage{multirow} 
\usepackage{amssymb}
\usepackage{subcaption}

\usepackage{hyperref}

\usepackage{graphicx, amsmath, amssymb, caption, subcaption, multirow, overpic, textpos}
\usepackage{colortbl}
\usepackage[american]{babel}

\definecolor{myyellow}{RGB}{156, 120, 54}
\definecolor{mygreen}{RGB}{143, 159, 121}
\definecolor{myblue}{RGB}{136, 149, 176}
\definecolor{skyblue}{RGB}{110, 150, 180}
\definecolor{nitrogenblue}{RGB}{80, 105, 246}
\definecolor{oxygenred}{RGB}{234, 63, 52}
\definecolor{sulfuryellow}{RGB}{244, 190, 81}
\definecolor{maskedgrey}{RGB}{239,239,239}
\definecolor{primaryblue}{RGB}{52, 152, 219}
\definecolor{accentorange}{RGB}{230, 126, 34}
\definecolor{warningred}{RGB}{231, 76, 60}
\definecolor{successgreen}{RGB}{46, 204, 113}
\definecolor{lightgray}{RGB}{236, 240, 241}

\definecolor{genesis-blue}{HTML}{0046FF}
\definecolor{genesis-integration}{HTML}{D8D8D8}
\definecolor{genesis-discovery}{HTML}{083651}
\definecolor{genesis-ai}{HTML}{00C885}
\definecolor{genesis-possibility}{HTML}{FFFFFF}
\definecolor{genesis-exploration}{HTML}{1D1F24}

\newcommand{\code}[1]{\texttt{#1}}
\newlength\savewidth\newcommand\shline{\noalign{\global\savewidth\arrayrulewidth
  \global\arrayrulewidth 1pt}\hline\noalign{\global\arrayrulewidth\savewidth}}
\newcommand{\tablestyle}[2]{\setlength{\tabcolsep}{#1}\renewcommand{\arraystretch}{#2}\centering\footnotesize}
\renewcommand{\paragraph}[1]{\vspace{1.25mm}\noindent\textbf{#1}}

\newcolumntype{x}[1]{>{\centering\arraybackslash}p{#1pt}}
\newcolumntype{y}[1]{>{\raggedright\arraybackslash}p{#1pt}}
\newcolumntype{z}[1]{>{\raggedleft\arraybackslash}p{#1pt}}
\newcommand{\app}{\raise.17ex\hbox{$\scriptstyle\sim$}}

\definecolor{deemph}{gray}{0.6}

\definecolor{baselinecolor}{gray}{.9}
\newcommand{\baseline}[1]{\cellcolor{baselinecolor}{#1}}
\definecolor{lightcolor}{RGB}{137, 207, 240}

\newcommand{\remove}[1]{}

\newcommand{\reffig}[1]{Figure~\ref{fig:#1}}

\newcommand{\refsec}[1]{Section~\ref{sec:#1}}

\newcommand{\reftbl}[1]{Table~\ref{tbl:#1}}
\newcommand{\refalg}[1]{Algorithm~\ref{alg:#1}}

\newcommand{\lblfig}[1]{\label{fig:#1}}
\newcommand{\lblsec}[1]{\label{sec:#1}}

\newcommand{\lbltbl}[1]{\label{tbl:#1}}
\newcommand{\lblalg}[1]{\label{alg:#1}}

\makeatletter
\DeclareRobustCommand\onedot{\futurelet\@let@token\@onedot}
\def\@onedot{\ifx\@let@token.\else.\null\fi\xspace}

\def\name{{\textit{Pairmixer}}\xspace}

\def\af{{\textit{AlphaFold3}}\xspace}

\title{Triangle Multiplication is All You Need for Biomolecular Structure Representations}

\author{
\hspace{-0.5em} Jeffrey Ouyang-Zhang\textsuperscript{1,2}
\thanks{Work done during an internship at Genesis Research \\ 
\texttt{\{jozhang,danny.diaz,klivans,philkr\}@cs.utexas.edu} \\
\texttt{\{pranav,gianscarpe,richard,ngruver,sandra,maruan\}@genesistherapeutics.ai}}
, Pranav Murugan\textsuperscript{1}, Daniel J. Diaz\textsuperscript{2}, Gianluca Scarpellini\textsuperscript{1},
\AND Richard Strong Bowen\textsuperscript{1}, Nate Gruver\textsuperscript{1},
Adam Klivans\textsuperscript{2}, Philipp Kr\"ahenb\"uhl\textsuperscript{2},
\AND Aleksandra Faust\textsuperscript{1}, Maruan Al-Shedivat\textsuperscript{1} \\
\textsuperscript{1}Genesis Research \quad \textsuperscript{2}UT Austin \\
}

\iclrfinalcopy %
\begin{document}

\maketitle

\begin{abstract}
AlphaFold has transformed protein structure prediction, but emerging applications such as virtual ligand screening, proteome-wide folding, and de novo binder design demand predictions at a massive scale, where runtime and memory costs become prohibitive.
A major bottleneck lies in the Pairformer backbone of AlphaFold3-style models, which relies on computationally expensive triangular primitives—especially triangle attention—for pairwise reasoning.
We introduce \name, a streamlined alternative that eliminates triangle attention while preserving higher-order geometric reasoning capabilities that are critical for structure prediction.
\name substantially improves computational efficiency, matching state-of-the-art structure predictors across folding and docking benchmarks, delivering up to $4\times$ faster inference on long sequences while reducing training cost by 34\%.
Its efficiency alleviates the computational burden of downstream applications such as modeling large protein complexes, high-throughput ligand and binder screening, and hallucination-based design.
Within BoltzDesign, for example, \name delivers over $2\times$ faster sampling and scales to sequences $\sim$30\% longer than the memory limits of Pairformer. 
Code is available at \href{https://github.com/genesistherapeutics/pairmixer}{https://github.com/genesistherapeutics/pairmixer}.

\end{abstract}

\section{Introduction}

AlphaFold~\citep{senior2020improved,jumper2021alphafold2} has transformed protein structure prediction and become an indispensable tool across the biological sciences. Yet emerging applications increasingly demand massive scale. Virtual screening of protein–ligand interactions, modeling of large protein complexes, proteome-wide folding, and iterative de novo binder design already require millions (and soon billions) of inference calls. At this scale, runtime and memory efficiency are critical bottlenecks: for example, Boltz-1~\citep{wohlwend2024boltz1} requires over 15 minutes to process a single 2048-token sequence on an A100 GPU (see~\refsec{inference_speed}).
The dominant computational cost comes from pairwise token representations and triangular primitives, which scale \emph{cubically} with sequence length $L$. 
While triangle multiplication is implemented efficiently via matrix multiplications, triangle attention requires $L$ attention operations, introducing substantial memory and runtime overhead.

\begin{figure}[t]
\centering
\includegraphics[width=0.99\linewidth]{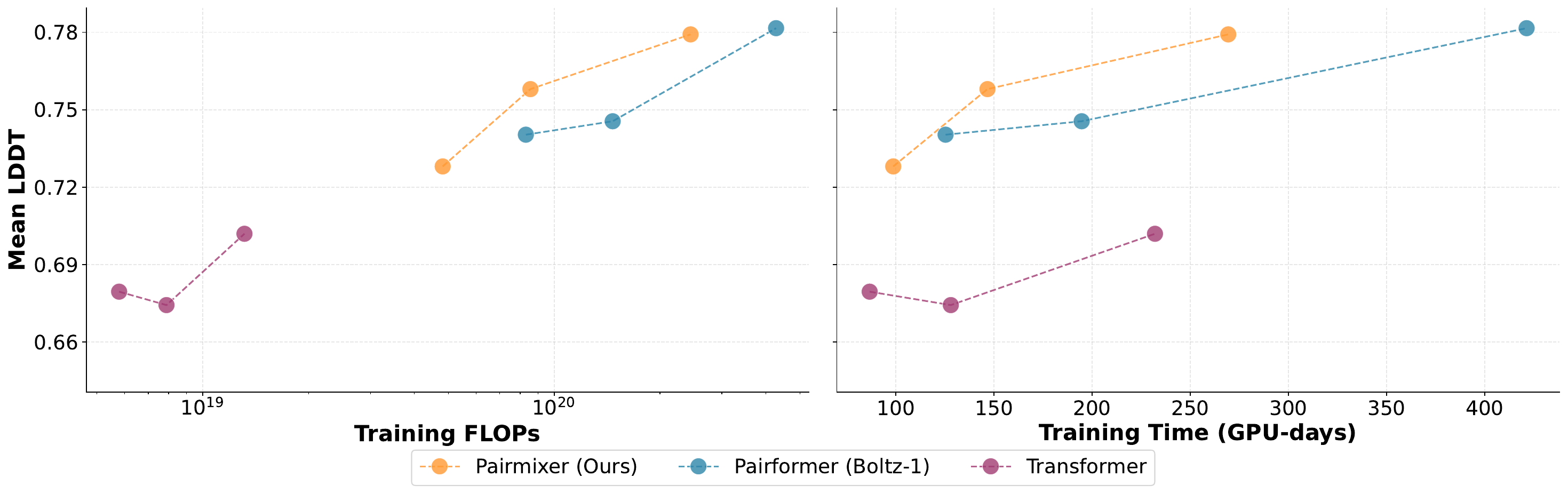}
\caption{
\textbf{
\name is an efficient architecture for biomolecular structure prediction.
}
Across multiple model sizes, \textcolor{orange}{\name} matches the performance of the leading \textcolor{MidnightBlue}{Pairformer} architecture while delivering greater training efficiency.
}
\vspace{-15pt}
\label{fig:pull}
\end{figure}

We introduce \name, a streamlined alternative to the Pairformer backbone of AlphaFold3~\citep{abramson2024alphafold3}.
By retaining triangle multiplication and feed-forward networks while eliminating triangle and sequence attentions, \name preserves the ability to reason over higher-order geometric interactions that are critical for structure prediction while alleviating Pairformer's heavy computational burden.
Despite this simplification, \name performs comparably on RCSB and CASP15 test sets against state-of-the-art predictors such as AlphaFold, Chai-1, and Boltz-1, while providing $4\times$ faster inference on long sequences.
\name consistently matches the performance of Pairformer backbone across protein-ligand, antibody-antigen, protein-nucleic acid and RNA structures while training in 34\% fewer GPU-days across multiple model sizes (see~\reffig{pull}).

By reducing both runtime and memory requirements, \name expands the scope of feasible downstream applications of structure prediction.
It enables modeling of larger protein complexes beyond the limits of triangle attention, supports high-throughput screening of ligands and binders, and accelerates hallucination-based design pipelines~\citep{pacesa2025bindcraft}. Within the BoltzDesign1~\citep{cho2025boltzdesign1} framework, \name provides over $2\times$ faster sampling and scales to sequences beyond 700 amino acids, where BoltzDesign otherwise fails due to memory overflow.
Our analysis highlights the role of the pair representation in learning precise distances between residues and suggests that triangle multiplication learns to capture sparse long-range interactions among residue triplets.

\section{Related Work}
\vspace{-1em}

\paragraph{Biomolecular Structure Prediction. }
Protein structure prediction has progressed rapidly in recent years, with early efforts primarily focused on modeling monomeric proteins~\citep{senior2020improved,jumper2021alphafold2,baek2021rosettafold,yang2020improved,ahdritz2024openfold}.
As these approaches matured, structure predictors expanded to handle multimeric assemblies~\citep{evans2021protein,baek2023rosettafold2} and other modalities such as nucleic acids~\citep{baek2024accurate}.
Today, state-of-the-art predictors can fold complexes that span a wide range of biomolecular types~\citep{abramson2024alphafold3,team2025intfold,boitreaud2024chai,wohlwend2024boltz1,bytedance2025protenix}.

Biomolecular structure predictors rely on specialized backbones that capture complex geometric relationships among molecular entities.
Early approaches such as trRosetta~\citep{yang2020improved} and AlphaFold1~\citep{senior2020improved} leveraged convolutional neural networks to extract pairwise residue features from multiple sequence alignments (MSAs) and predict inter-residue distances.
AlphaFold2 introduced the transformer-based Evoformer to jointly model MSA and pair representations~\citep{jumper2021alphafold2}, while AlphaFold3 refined it with the Pairformer, which decouples MSA and pair processing~\citep{abramson2024alphafold3}.
The Pairformer has since become the de-facto backbone architecture for biomolecular structure prediction~\citep{team2025intfold,boitreaud2024chai,wohlwend2024boltz1,bytedance2025protenix}.
However, despite its strong performance, the Pairformer remains complex and computationally demanding.

Several alternative architectures have been proposed to simplify structure prediction backbones.
MiniFold~\citep{wohlwendminifold} streamlines Alphafold2's Evoformer using a lightweight Miniformer based on triangle multiplications.
SimpleFold~\citep{wang2025simplefold} replaces the Evoformer with a sequence-only transformer that omits pair representations.
Our work also simplifies backbone design, but unlike prior efforts focused on monomeric folding, \name is developed for AlphaFold3-like cofolding models, enabling structure prediction across broader biomolecular modalities.

\paragraph{Downstream Applications of Structure Prediction. }
The success of biomolecular structure prediction has enabled a growing number of downstream applications, many of which leverage predicted structures at unprecedented scales.
Large-scale resources such as the AlphaFold Database~\citep{varadi2022alphafold} and OpenFold~\citep{ahdritz2024openfold} have generated massive synthetic protein structure datasets using AlphaFold2, powering advances in structure search~\citep{van2024foldseek}, protein language modeling~\citep{prott5,ouyang2024distilling,hayes2025simulating}, and diffusion-based structure generation~\citep{geffner2025proteina,lin2024genie2,daras2025ambient}.
Structure predictors now drive virtual screening pipelines that evaluate millions of candidate drugs based on predicted protein–ligand interactions~\citep{wong2022benchmarking,shamir2025state,scardino2023good}, and large-scale folding studies that map the human interactome~\citep{ille2025human,zhang2025predicting}.
Hallucination-based generation methods such as BindCraft~\citep{pacesa2025bindcraft} further use predictors in iterative optimization loops requiring millions of model evaluations.
As these applications expand in scope and scale, inference speed becomes a critical bottleneck.
We introduce a structure predictor that matches state-of-the-art accuracy while operating at a fraction of the runtime, enabling faster and broader deployment of downstream workflows.

\paragraph{Attention-free Architectures. }
While transformers lead modern architectures, attention-free variants aim to improve scalability.
FNet~\citep{lee2021fnet} and related models~\citep{poli2023hyena,zhai2021attnfree} replace attention with Fourier or convolutional mixing for sub-quadratic efficiency, while MLP-Mixer~\citep{tolstikhin2021mlp} achieves competitive performance using token- and channel-wise multi-layer perceptrons (MLPs).
\name removes attention entirely from the backbone and mixes tokens through matrix multiplication.

Architectures based on triangle multiplication have been explored in several prior works.
Genie2~\citep{lin2024genie2} performs de-novo structure generation by iteratively updating a pair representation through triangle multiplications, while MSA Pairformer~\citep{akiyama2025scaling} applies similar operations to extract features from multiple sequence alignments.
IgFold~\citep{ruffolo2023fast} incorporates triangle operations within GNN layers.
\name likewise learns rich protein representations through triangle multiplication, but in the context of biomolecular structure prediction.

\begin{figure}[t]
\centering
\vspace{-10pt}
\includegraphics[width=0.99\linewidth]{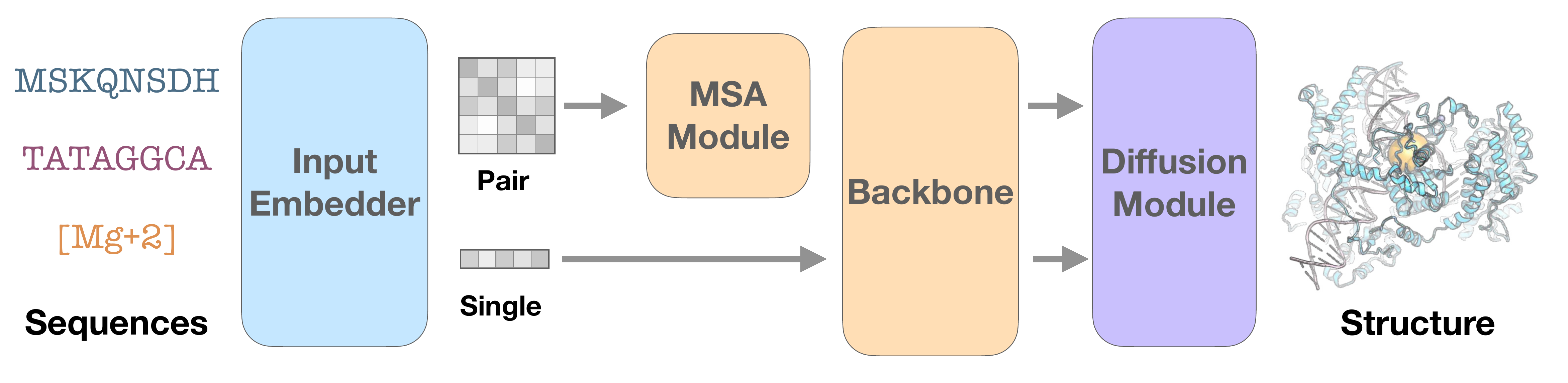}
\caption{
\textbf{Overview of Biomolecular Structure Prediction.}
Given a list of sequences, our model predicts the 3D folded structure of all sequences within a single complex.
Input sequences are first \textit{\textcolor{MidnightBlue}{embedded}} into a single representation for each residue and a pair representation to capture the relationship between pairs of residues.
The \textit{\textcolor{orange}{MSA Module}} and \textit{\textcolor{orange}{Backbone}} (e.g., Pairformer) extracts deep pairwise features capturing inter-residue interactions, which are then passed to the \textit{\textcolor{Mulberry}{diffusion module}} to generate the 3D structure.
(Additional inputs such as MSAs, conformers, and templates are omitted for clarity.)
}
\lblfig{pipeline}
\vspace{-15pt}
\end{figure}

\section{Preliminaries}
\lblsec{preliminaries}
Let $x = \{ x^{(1)}, \cdots, x^{(K)} \}$ denote a collection of $K$ biomolecular sequences.
Each sequence $x^{(k)}=(x^{(k)}_1, \cdots, x^{(k)}_{L^{(k)}}) $ consists of tokens $x^{(k)}_i \in \mathcal{T}$ %
corresponding to an amino acid%
, a nucleic acid%
, or small molecule heavy atoms.
$L^{(k)}$ denotes the number of tokens in biomolecule $x^{(k)}$.
The goal of biomolecular structure prediction is to map the sequences $x$ to a three-dimensional structure $ a = \{ a^{(1)}, \cdots, a^{(K)} \}, $ where each biomolecular structure $ a^{(k)} = (\va^{(k)}_1, \cdots, \va^{(k)}_{N^{(k)}}) $ consists of atomic coordinates $\va^{(k)}_j \in \mathbb{R}^3$, and $N^{(k)}$ denotes the number of atoms in biomolecule $k$.
See \reffig{pipeline} for an overview.

\paragraph{The Input Embedder} concatenates the sequences $x = \{ x^{(1)}, \ldots, x^{(K)} \}$ and embeds it into a ``\textit{single}'' length $L = \sum_{k=1}^K L^{(k)}$ sequence representation $\vs^{\text{init}} \in \mathbb{R}^{L \times C_s}$ of dimension $C_s$.
Modern structure predictors~\citep{jumper2021alphafold2} additionally initialize a ``\textit{pair}'' representation $\vz^{\text{init}} \in \mathbb{R}^{L \times L \times C_z}$:
$$
\vz_{ij} = \vs_i + \vs_j + \mathbf{PE}(i, j),
$$
where $\mathbf{PE}(i, j)$ is a positional encoding that incorporates both intra- and inter-sequence distances and $C_z$ is the pair embedding dimension.
Intuitively, $\vz_{ij} \in \mathbb{R}^{C_z}$ captures the relational context between tokens $\vs_i$ and $\vs_j$ and enables the model to reason about longer-range couplings.
Since pairwise reasoning is critical for structure prediction, we adopt the same input embedding scheme.

\begin{figure}[t]
    \centering
    \begin{subfigure}[t]{\linewidth}
        \centering
        \includegraphics[width=\linewidth]{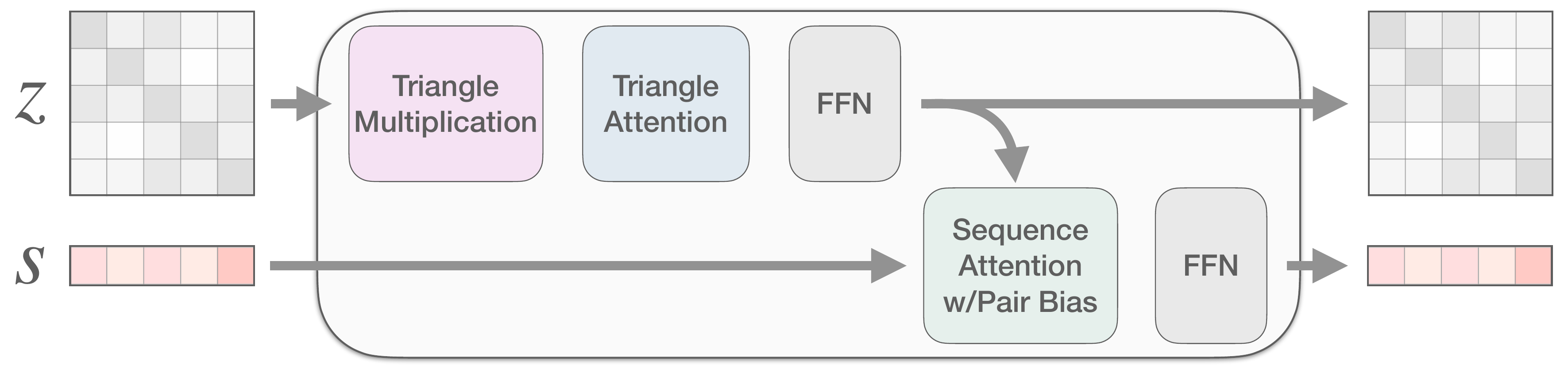}
        \caption{
        \textbf{Pairformer architecture.}
        The de facto biomolecular structure prediction backbone.
        }
        \label{fig:pairformer}
    \end{subfigure}
    \vspace{1em}
    \begin{subfigure}[t]{\linewidth}
        \centering
        \includegraphics[width=\linewidth]{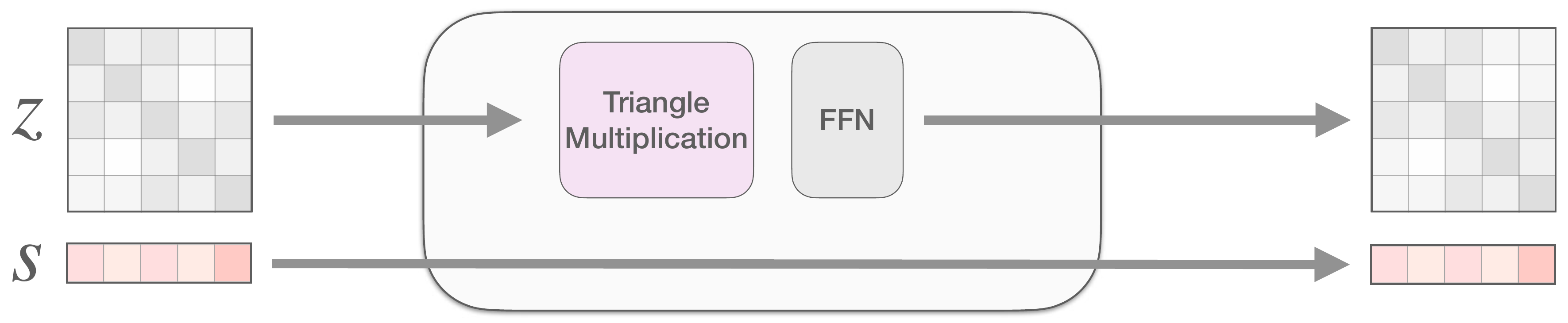}
        \caption{
        \textbf{\name architecture.}
        An efficient yet effective biomolecular structure prediction backbone.
        }
        \label{fig:pairmixer}
    \end{subfigure}
    \vspace{-1.5em}
    \caption{
        \textbf{Schematic comparison of the Pairformer and \name backbones.}
        \name simplifies the Pairformer architecture by removing redundancies.
        This results in faster training and inference, expanding the scale of downstream applications.
    }
    \vspace{-2em}
    \label{fig:pairmixer_vs_pairformer}
\end{figure}

\paragraph{The MSA Module} encodes evolutionary information that is crucial for structure prediction    ~\citep{benner1991patterns,yanofsky1964protein,ovchinnikov2017protein,ovchinnikov2014robust,morcos2011direct,weigt2009identification}.
For each amino acid or nucleic acid sequence $x^{(k)}$, we perform a homology search to construct a multiple sequence alignment (MSA) of related sequences that likely adopt the same fold.
Formally, $\mathbf{MSA}(x^{(k)}) \in \left( \mathcal{T} \cup \{\code{GAP}\} \right) ^{M^{(K)} \times L^{(K)}}$ contains $M^{(k)}$ aligned sequences of length $L^{(k)}$.
This alignment establishes positional correspondence across homologous sequences, enabling detection of conserved sites and co-evolutionary couplings.
The resulting MSAs are then paired, concatenated, and embedded into $\vm^{\text{init}} \in \mathbb{R}^{M \times L \times C_m}$ where $M$ is the number of filtered homologous sequences and $C_m$ is the MSA embedding dimension.

The MSA module takes $(\vm^{\text{init}}, \vz^{\text{init}})$ as input, extracts structurally-relevant evolutionary patterns from $\vm^{\text{init}}$, and encodes pairwise interactions into $\vz^{\text{msa}}$ to guide folding.
Since processing all $M$ sequences in the MSA is computationally expensive, AlphaFold3 introduced a shallow 4-layer MSA module after which the MSA is discarded while the evolutionary-aware pair representation $\vz^{\text{msa}}$ continues to be refined.
Our model derives $\vz^{\text{msa}}$ from an MSA module but introduces a more efficient feature extractor to refine its evolutionary signals.

\paragraph{The Pairformer backbone} serves as the primary feature extractor for AlphaFold3~\citep{abramson2024alphafold3}, producing structrually-aware representations that encode geometric constraints between residues (see~\reffig{pairformer}).
It takes $(\vs^{\text{init}}, \vz^{\text{msa}})$ as input and employs several specialized modules that iteratively update the sequence and pair representations to produce $(\vs^{\text{backbone}}, \vz^{\text{backbone}})$.
See~\reffig{pairformer_combined} for a more detailed treatment of the entire architecture.

The Pairformer contains two specialized modules for processing the pair representation: triangle attention and triangle multiplication.
These modules treat the pair representation $\vz \in \mathbb{R}^{L \times L \times C_z}$ as edge features of a fully-connected graph of $L$ nodes and reason over triplets of residues (nodes) to learn geometric constraints.

\textit{Triangle attention} computes attention (with pair bias) along every row (and column) of the pair representation.
Formally, the update to row $i$ is
$$\mathbf{TriAtt}(\vz)_{i} = \text{softmax} \Big( (\mW_Q \vz_i)(\mW_K \vz_i)^\top + \mW_B \vz \Big)\, \mW_V \vz_i$$
where $(\mW_Q, \mW_K, \mW_V)$ are standard attention projection matrices, and $\mW_B$ projects the pair representation into an attention bias term~\footnote{single head and removed scaling for brevity}.
$\mathbf{TriAtt}(\vz)_{i}$ effectively performs attention over all residues while conditioning on residue $i$.

\textit{Triangle multiplication} performs matrix multiplications to integrate features across different rows (and columns) of the pair representation.
Formally, the update to edge $\vz_{ij}$ is 
$$\mathbf{TriMul}(\vz)_{ij} = \sum_{k=1}^L ( \mW_a \vz_{ik}) \odot (\mW_b \vz_{jk})$$
where $\mW_a, \mW_b$ are linear projection layers.
For each edge $\vz_{ij}$, triangle multiplication computes how every node $k$ interacts with query nodes $i$ and $j$ through edges $\vz_{ik}$ and $\vz_{jk}$.

Both operations scale cubically with sequence length, making the processing of long sequences computationally expensive. 
Triangle multiplication is more efficient, as it can be implemented with matrix multiplications (e.g., \code{torch.einsum}), 
whereas triangle attention incurs the higher cost of $L$ full attention computations. 
In this work, we streamline the cofolding backbone to its essential components and show that triangle multiplication yields representations as powerful as those from triangle attention, but at substantially lower computational cost, supporting a range of downstream applications.

While the Pairformer is trained with an auxiliary distogram loss that ensures $\vz^{\text{backbone}}$ accurately represents all pairwise token distances, it does not yet specify an atomic 3-D structure.

\paragraph{The Diffusion Module} samples the atomic coordinates conditioned on $(\vs^{\text{backbone}}, \vz^{\text{backbone}})$.
It uses transformers to derive atomic representations from the token-level sequence and pair representations, and subsequently denoises all-atom coordinates based on these representations.
We leverage the diffusion module as-is to realizes 3-D structures conditioned on single and pair representations derived from our efficient backbone.

\begin{algorithm}[t]
\caption{\name Backbone}
\lblalg{pairmixer}
\begin{algorithmic}[1]
\REQUIRE Input pair representation $\vz^{\text{msa}} \in \mathbb{R}^{L \times L \times C_z}$
\REQUIRE Number of backbone layers $N$
\ENSURE Updated pair representation $\vz_N$
\STATE $\vz_0 \leftarrow \vz^{\text{msa}}$
\FOR{$l = 0$ to $N-1$}
    \STATE $\vz_l \leftarrow \vz_l + \mathbf{TriMulIncoming}(\vz_l)$
    \STATE $\vz_l \leftarrow \vz_l + \mathbf{TriMulOutgoing}(\vz_l)$
    \STATE $\vz_{l+1} \leftarrow \vz_l + \mathbf{FFN}(\vz_l)$
\ENDFOR
\RETURN $\vz_N$
\end{algorithmic}
\end{algorithm}

\section{Method}
We introduce \name, an attention-free feature extractor for biomolecular structure prediction and design (see~\reffig{pairmixer_vs_pairformer}).
\name exclusively updates the pair representation $\vz^{\text{msa}}$, leaving the single-sequence representation $\vs^{\text{init}}$ unchanged.
Through \textit{triangle multiplication}, \name efficiently mixes features within the pair representation, facilitating reasoning over residue triplets and their geometric constraints.
Combined with feed-forward networks (FFN) that process all residue pairs, this architecture provides an effective and expressive backbone for biomolecular structure prediction.

The full algorithmic specification of \name is available in \refalg{pairmixer}.
In developing \name, we identified and removed two unnecessary modules from the Pairformer: sequence updates and triangle attention.

\paragraph{Removing Sequence Updates.}
In AlphaFold2's Evoformer backbone, sequence updates were essential components that processed the MSA to capture evolutionary features.
However, the MSA Module in cofolding models now preprocesses the MSA and encodes this evolutionary information directly into the pair representation $\vz^{\text{msa}}$, eliminating the need for sequence updates to provide evolutionary information.
Since the pair updates proved more expressive, we bypass sequence processing entirely and pass the initial sequence representation directly to the diffusion module (i.e., $\vs^{\text{backbone}} = \vs^{\text{init}}$).

\paragraph{Removing Triangle Attention.}
Triangle attention reasons over residue triplets by applying attention to each row of the pair representation $\vz_i$, using the full $\vz$ as pairwise bias (see~\reffig{triangle_attention_arch}).
However, this approach is computationally expensive, requiring $L$ separate attention operations over $L$ tokens per layer.
Triangle multiplication offers equivalent capability for capturing geometrically consistent pair representations via a triplet reasoning mechanism, but with significantly lower computational cost.
Since both methods have independently demonstrated strong performance in structure prediction~\citep{jumper2021alphafold2}, we adopt the more efficient triangle multiplication approach.

\section{Results}

\subsection{Implementation Details}

We implement \name on top of Boltz-1, an \af descendant.
More specifically, we replace the Pairformer backbone with \name and remove triangle attention from the MSA Module.
Note that we do not alter the diffusion module's transformer architecture.
We also introduce a transformer baseline that preserves the sequence update while removing the pair update in the backbone.
To ensure this baseline is as strong as possible, we modify the architecture to allow features to flow effectively from the MSA module into the diffusion module (see~\refsec{transformer_baseline}).

Following Boltz-1 training schedule~\citep{wohlwend2024boltz1}, we train on 384/3456 token/atom crops for the first 53k iterations using the PDB and OpenFold distillation dataset.
We then finetune for 15k iterations on the PDB dataset with a larger crop size of 512/4608.
To evaluate the generality of our approach, we train models of multiple sizes.
Our large configuration matches Boltz-1, with 48 Pairformer layers and 24 diffusion transformer layers.
In addition, we develop small and medium variants with 12/24 Pairformer layers and 6/24 diffusion transformer layers, respectively.
During inference, we default to 10 recycling steps and 200 sampling steps for all models.
In our main evaluation, we sample 5 poses and report the metrics on the top pose (oracle evaluation).
Full hyperparameter details are in~\reftbl{boltz1_hyperparameters}.

\begin{figure}[t]
\centering
\vspace{-10pt}
\includegraphics[width=0.99\linewidth]{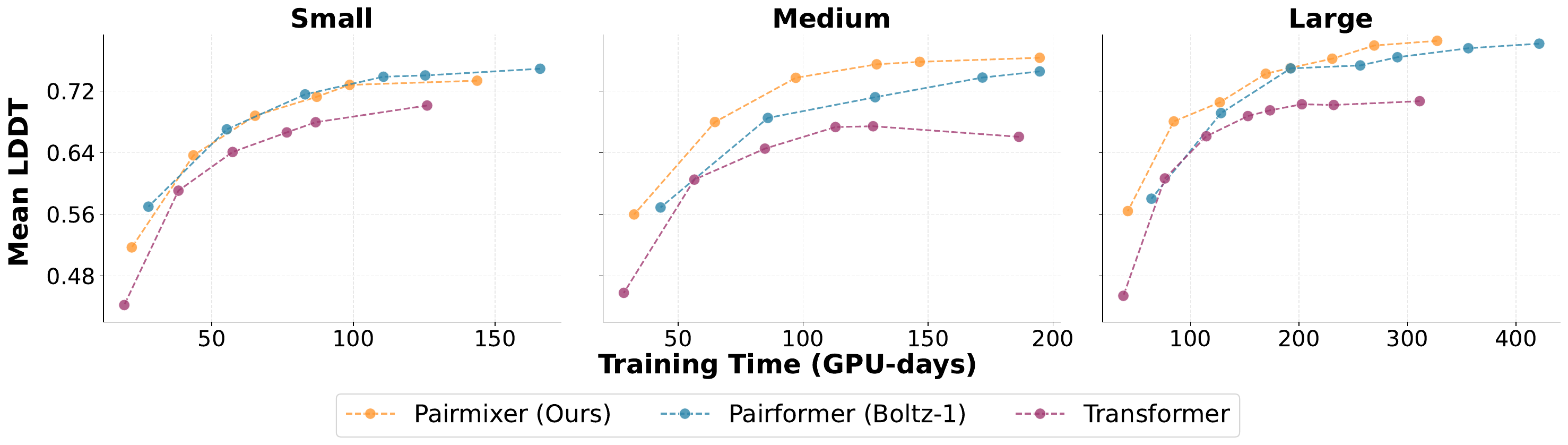}
\caption{
\textbf{Performance curves on RCSB test set across model sizes.}
We compare three backbone architectures across three model sizes over training.
\textcolor{orange}{\name} matches or surpasses the \textcolor{MidnightBlue}{Pairformer} baseline while training more efficiently.
}
\label{fig:results_test_rcsb_lddt_allsize}
\vspace{-1pt}
\end{figure}

\subsection{Comparisons on Cofolding performance across model sizes}
\lblsec{main_results}

We evaluate our efficient \name architecture against two baselines, Pairformer~\citep{abramson2024alphafold3} and a sequence-only Transformer.
All models are evaluated on the RCSB test set introduced in Boltz-1~\citep{wohlwend2024boltz1}, which contains 533 structures with at most 40\% sequence identity to the training set, maximum small-molecule similarity of 80\%, and resolution better than 4.5Å.
All models are evaluated at 15, 30, 45, 60, and 68 epochs, totalling 53k iterations and the large model is additionally evaluated during the second phase of 15k iterations.
We additionally extend training for small and medium \name and Transformer models until the total training time matches the Pairformer.
We report the final mean LDDT, averaged across all residues.

Our \name consistently outperforms or matches the Pairformer across all model sizes (see~\reffig{results_test_rcsb_lddt_allsize}).
At the large scale, \name reaches Pairformer-level accuracy (mean LDDT of 0.78) while requiring only 66\% of the training time.
The trend holds at smaller scales: \name surpasses Pairformer at the medium scale and matches it at the small scale under equal training budgets.
Furthermore, under the same training time, \name exceeds the sequence-only Transformer baseline across all scales.
These results suggest that a sequence-only Transformer is inadequate for extracting structural features, while the triangle multiplications and feed-forward networks in \name are sufficient to capture rich structural representations.
Full tabular results are provided in~\reftbl{full_results_rcsb} and~\reftbl{full_results_casp15}, and detailed FLOPs analysis is provided in~\refsec{appendix/flops}.

\subsection{Inference time comparisons}
\lblsec{inference_speed}

Many downstream applications require running the structure predictor on thousands to millions of complexes, making inference efficiency critical.
In~\reffig{results_test_rcsb_lddt_allsize}, we benchmark \name against the Pairformer and a sequence-only transformer under a default setting of 512 tokens, 4608 atoms, MSA depth of 4096, 10 recycles, 48 blocks, and 200 sampling steps.

On this setup, Boltz-1 requires 34 seconds to generate a single sample on a GH200 GPU, while \name completes in 21 seconds, yielding a 1.6$\times$ speedup.
This advantage holds consistently across different recycle counts, MSA depths, and backbone sizes.
The scaling benefits are even more striking for longer sequences: at 1024 tokens, \name is 2$\times$ faster, and at 2048 tokens, it delivers a 4$\times$ speedup, reducing runtime from 1000 seconds to 250 seconds.
These results establish \name as a scalable and efficient architecture, making large-scale cofolding more practical.

\begin{figure}[t]
\centering
\vspace{-10pt}
\includegraphics[width=0.99\linewidth]{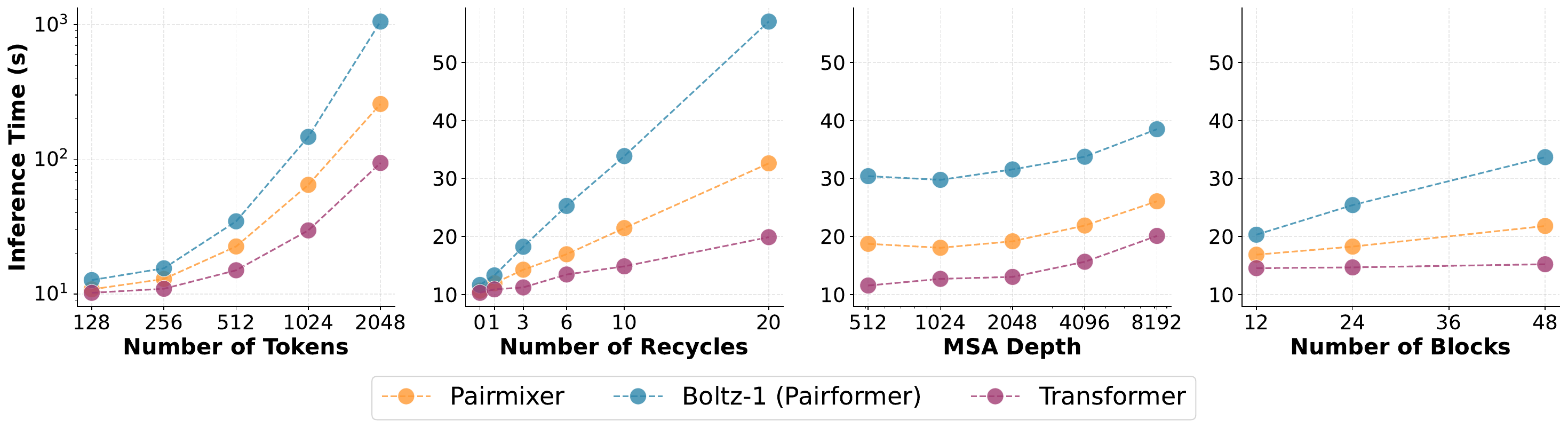}
\caption{
\textbf{Inference speed analysis.}
We measure runtime across architectures and input sizes.
While the \textcolor{violet}{Transformer} is the fastest overall, \textcolor{orange}{\name} achieves substantially lower inference times than \textcolor{MidnightBlue}{Pairformer}, particularly on longer sequences.
}
\label{fig:inference_time_scaling}
\vspace{-1pt}
\end{figure}

\begin{figure}[b]
\centering
\vspace{-10pt}
\includegraphics[width=0.99\linewidth]{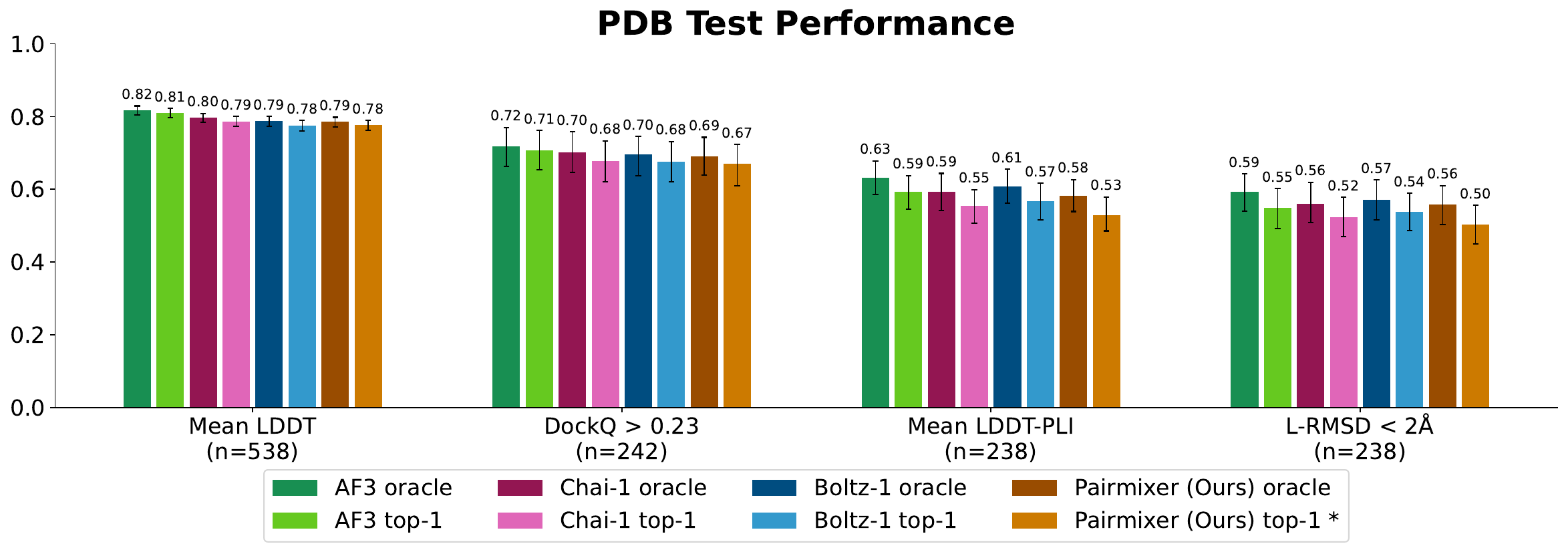}
\caption{
\textbf{System-level comparison on the RCSB test set.}
We evaluate against AlphaFold3, Chai-1, and Boltz-1 on protein and small-molecule structure prediction. 
\textcolor{brown}{\name} performs competitively with these state-of-the-art approaches. 
Error bars denote bootstrapped 95\% confidence intervals.
$^*$Since we do not train a confidence model, results are reported using the first prediction. 
}
\label{fig:comparisons_against_boltz1}
\vspace{-1pt}

\end{figure}

\subsection{Comparisons to prior works}

\reffig{comparisons_against_boltz1} compares \name to other cofolding models on the RCSB test set, evaluating protein folding, protein–protein interactions (DockQ), and protein–ligand interactions (lDDT-PLI and ligand RMSD < 2).
See~\refsec{main_results} for a description of the test dataset.
We generate five poses per complex and report both the performance of the best pose (oracle) and the average across poses.
Results for existing methods are taken from the literature.
\name matches Boltz-1 in mean lDDT and protein–ligand lDDT, slightly improves ligand RMSD < 2 (0.55 vs. 0.54), but lags on DockQ > 0.23 (0.63 vs. 0.64).
These results indicate that even at the largest scale, triangle multiplication and pair FFNs in \name are sufficient for cofolding across diverse interaction types.
We show similar results on the CASP15 test set in~\refsec{casp_results}.

\subsection{Comparisons on diverse structure prediction tasks}

\reftbl{more_benchmarks} shows evaluation results across a variety of biomolecular structure prediction benchmarks, including protein–ligand complexes (PoseBusters), antibody–antigen complexes, protein–nucleic acid complexes, and RNA structures.
Experimental details are provided in~\refsec{diverse_task_details}.
Pairmixer performs on par with Pairformer, while standard Transformers generally lag behind.
The exception is RNA structures, where the Transformer baseline slightly outperforms both Pairformer and Pairmixer, likely due to the limited availability of RNA structural training data.
Notably, Pairmixer achieves comparable performance to Pairformer despite not using sequence attention.
These results highlight Pairmixer's generality and robustness in modeling diverse biomolecular interactions.

\begin{table}[t]
\centering
\caption{
\textbf{Performance on diverse biomolecular structure prediction benchmarks.} 
Number of test samples for each dataset is indicated in parentheses.
}
\lbltbl{more_benchmarks}
\vspace{-5pt}
\begin{subtable}[t]{0.48\linewidth}
\caption{\textbf{PoseBusters}: Protein-Ligand Complex (298)}
\vspace{-5pt}
\tablestyle{5pt}{1.05}
\begin{tabular}{lccc}
Method & $\text{RMSD}_{<2}$ & $\text{RMSD}_{<1}$ & $\text{lDDT}_\text{PLI}$ \\
\shline
Pairformer (Boltz-1) & 0.68 & 0.46 & 0.74 \\
Pairmixer (Ours) & 0.67 & 0.45 & 0.73 \\
\end{tabular}
\lbltbl{posebusters}
\end{subtable}
\hfill
\begin{subtable}[t]{0.48\linewidth}
\caption{\textbf{Antibody–Antigen Complex} (70)}
\vspace{-5pt}
\tablestyle{5pt}{1.05}
\begin{tabular}{lc}
Method & $\text{DOCKQ}_{>0.23}$ \\
\shline
Pairformer (Boltz-1) & 0.23 \\
Pairmixer (Ours) & 0.23 \\
Transformer & 0.08 \\
\end{tabular}
\lbltbl{antibodies}
\end{subtable}

\vspace{6pt}

\begin{subtable}[t]{0.48\linewidth}
\caption{\textbf{Protein–Nucleic Acid Complex} (172)}
\vspace{-5pt}
\tablestyle{5pt}{1.05}
\begin{tabular}{lcc}
Method & ICS & IPS \\
\shline
Pairformer (Boltz-1) & 0.50 & 0.65 \\
Pairmixer (Ours) & 0.51 & 0.66 \\
Transformer & 0.48 & 0.64 \\
\end{tabular}
\lbltbl{protein_nucleic_acids}
\end{subtable}
\hfill
\begin{subtable}[t]{0.48\linewidth}
\caption{\textbf{RNA Structure} (27)}
\vspace{-5pt}
\tablestyle{5pt}{1.05}
\begin{tabular}{lc}
Method & lDDT \\
\shline
Pairformer (Boltz-1) & 0.58 \\
Pairmixer (Ours) & 0.59 \\
Transformer & 0.61 \\
\end{tabular}
\lbltbl{rna_only}
\end{subtable}
\vspace{-5pt}
\end{table}

\subsection{Comparisons on binder design (BindFast)}

Hallucination-based protein design methods have shown that structure predictors can act as differentiable scoring functions for sequence optimization.
However, they are memory-intensive and slow, requiring hundreds of runs to generate a single sequence.
We introduce BindFast, which replaces BoltzDesign's~\citep{cho2025boltzdesign1} Pairformer backbone with \name, reducing runtime and memory usage.
On 80GB A100 GPU, BoltzDesign encountered OOM errors on targets with over 500 residues, while BindFast handled targets up to 650 residues (+30\%) and ran over 2$\times$ speedups (see~\reftbl{bindfast}).
Qualitative comparisons in~\reffig{bindfast_qual} show comparable designs, suggesting BindFast enables faster in-silico iteration and design of larger, biologically relevant binders.
Details are in~\refsec{bindfast_details}.

\begin{table}[b]
\centering
\caption{
\textbf{Runtime comparison of generating proteins with \name and Pairformer in the BoltzDesign framework.}
For biologically relevant targets of various sequence lengths, we generate three 110-residue binders using 160 iterations in all settings and report the average running time.
}
\vspace{-1em}
\setlength{\tabcolsep}{6pt}
\renewcommand{\arraystretch}{1.2}
\begin{tabular}{lcccccc}
Target & 
\begin{tabular}[c]{@{}c@{}}PDB\_Chain\end{tabular} &
\begin{tabular}[c]{@{}c@{}}Complex \\ Length\end{tabular} &
\begin{tabular}[c]{@{}c@{}}Target \\ Length\end{tabular} &
\begin{tabular}[c]{@{}c@{}}Pairformer \\ Time (sec)\end{tabular} &
\begin{tabular}[c]{@{}c@{}}Pairmixer \\ Time (sec)\end{tabular} &
\begin{tabular}[c]{@{}c@{}}Speedup \\ \end{tabular} \\
\shline
GIP peptide & 2QHK\_B & 140 & 30 & 680 & 337 & 2.01× \\
Ubiquitin & 1UBQ\_A & 186 & 76 & 1113 & 532 & 2.09× \\
TP53 & 4MZI\_A & 303 & 193 & 3198 & 1390 & 2.30× \\
hSDH & 1P5J\_A & 429 & 319 & 7289 & 2920 & 2.50× \\
hMAO & 1GOS\_A & 607 & 497 & 17134 & 6601 & 2.60× \\
\small{bsDNA Polymerase} & 3TAN\_A & 702 & 592 & OOM & 9184 & $\infty$ \\
hTLR3 & 1ZIW\_A & 739 & 629 & OOM & 10568 & $\infty$ \\
\small{Prostate Antigen} \tiny{(PSA)} & 1Z8L\_A & 805 & 695 & OOM & OOM & -- \\
\end{tabular}
\lbltbl{bindfast}
\end{table}

\section{Analysis}
Predicting biomolecular structure requires reasoning over the entire sequence to capture diverse interactions among residues.
We analyze how the architectural design of modern structure predictors facilitates such reasoning and how the simplified \name architecture achieves this.

\paragraph{Pair representations. }
A central challenge in biomolecular structure prediction is determining the strength of the interactions between all residue pairs.
This is difficult because folding involves nonlocal tertiary interactions in which residues distant in sequence often interact physically in three-dimensional space.
Modern structure predictors address this challenge with a pair representation.
Our results indicate that the pair representation enables the model to capture fine-grain spatial relationships between all residue pairs.

We compare the performance of \name, which incorporates pair representations, against our sequence-only Transformer baseline in \reffig{win_rates_vs_transformer}.
On the lDDT metric computed from pairwise distances, \name achieves higher scores in 93.7\% of test complexes.
In contrast, on the RMSD metric, which requires global structural alignment, the improvement is smaller (74.7\%).
These findings show that \textit{pair representations provide greater benefits for local, pairwise accuracy} over sequence attention, suggesting their effectiveness in capturing residue–residue interactions.

\begin{figure}[t]
\centering
\vspace{-10pt}
\includegraphics[width=0.99\linewidth]{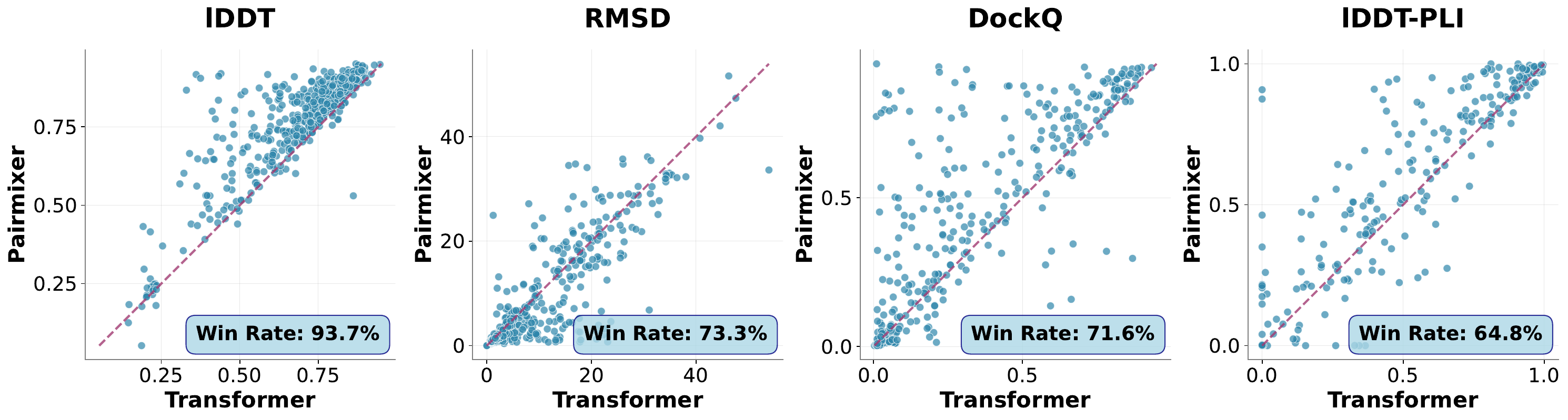}
\caption{
\textbf{Head-to-head comparison between \name and the Transformer backbone.}
The win rate shows how often the \name architecture achieves a better score than the Transformer architecture.
\name outperforms the Transformer on the distance-based lDDT metric in 93.7\% of the cases, highlighting that its advantage lies in capturing pairwise interactions.
}
\lblfig{win_rates_vs_transformer}
\vspace{-1pt}
\end{figure}

\begin{figure}[b]
\centering
\vspace{-10pt}
\includegraphics[width=0.99\linewidth]{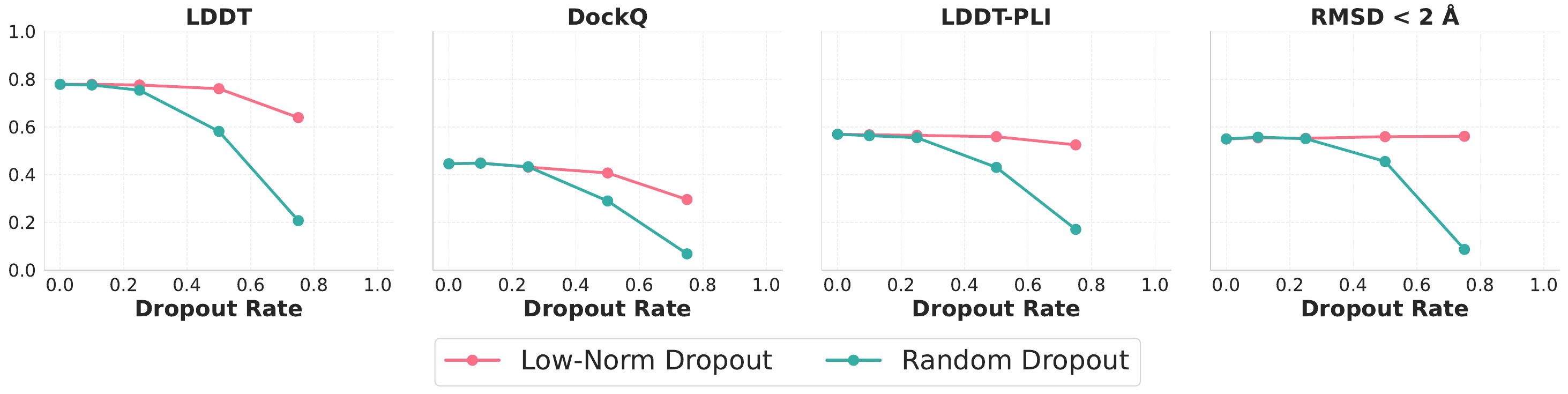}
\caption{
\textbf{\name Performance under different sparse triangle multiplication variants.}
The model is trained with standard triangle multiplication and evaluated under various dropout conditions.
While performance degrades rapidly under \textcolor{teal}{random dropout}, it remains stable when \textcolor{magenta}{low-norm} entries in the triangle multiplication are zeroed out.
}
\lblfig{sparsity}
\vspace{-1pt}
\end{figure}

\paragraph{Triangle multiplication. }
Modern structure predictors employ triangle attention and triangle multiplication within the pair representation to capture geometric relationships among residue triplets.
While triangle attention allows the model to reason \textit{sparsely} over interacting residues, triangle multiplication \textit{densely} aggregates features across the entire sequence.
However, our analysis shows that triangle multiplication also efficiently captures sparse geometric relationships among residue triplets by adjusting the magnitudes in the pair representations.

We explicitly sparsify triangle multiplication by introducing dropout during inference.
Formally,
\[
\mathbf{TriMulWithDropout}(\vz)_{ij} =
\sum_{k=1}^L (\mW_a \vz_{ik}) \odot (\mW_b \vz_{jk})
\, \cdot \underbrace{M(\vz_{ik}) \, M(\vz_{jk})}_{\text{new dropout masks}}
\]
where \(M(\vz_{ij}) \in \{0,1\}\) determines whether a particular interaction is active.

In \textit{random dropout} with dropout rate \(\gamma \in [0,1]\), the masks are sampled independently as
$M(\vz_{ik}), M(\vz_{jk}) \sim \mathrm{Bernoulli}(1-\gamma)$.
We experiment with a \textit{low-norm dropout} scheme, dropping any interaction $(i,j)$ whose pair representation lies in the \(\gamma \in [0,1]\) fraction of smallest magnitudes.
Formally, $M(\vz_{ik}) = \begin{cases} 1, & \text{if } k \in \text{Top}_{1-\gamma}(\{\|\vz_{il}\|\}_{l=1}^L) \\ 0, & \text{otherwise} \end{cases}$.
Under both dropout schemes, each term $(\mW_a \vz_{ik}) \odot (\mW_b \vz_{jk})$ is retained only if both corresponding masks $M(\vz_{ik})$ and $M(\vz_{jk})$ are active, resulting in a higher effective dropout rate.

\reffig{sparsity} shows the performance of the model where both dropout schemes are applied to every layer with $\gamma=0,0.10,0.25,0.50,0.75$.
We observe that performance starts to degrade rapidly once the random dropout rate exceeds 25\%, indicating that the model is not robust to random removal of interactions.
However, the performance is very similar under the low-norm dropout of 75\%.
This suggests that, like attention, triangle multiplication identifies and processes a small subset of interactions that are essential for accurate folding of biomolecular complexes.

To probe which interactions the model relies on in its sparse computation, we evaluate it using a local block-dropout scheme.
For block size $B$, we retain only local interactions: $M(\vz_{ik}) = \begin{cases} 1, & \text{if } |i-k| \le B \\ 0, & \text{otherwise}\end{cases}$.
The results in~\reffig{block} show that performance already begins to degrade $B=512$, with a substantial drop at $B=256$ tokens.
This suggests that \textit{triangle multiplication processes sparse, long-range interactions}.

\begin{figure}[t]
\centering
\vspace{0pt}
\includegraphics[width=0.99\linewidth]{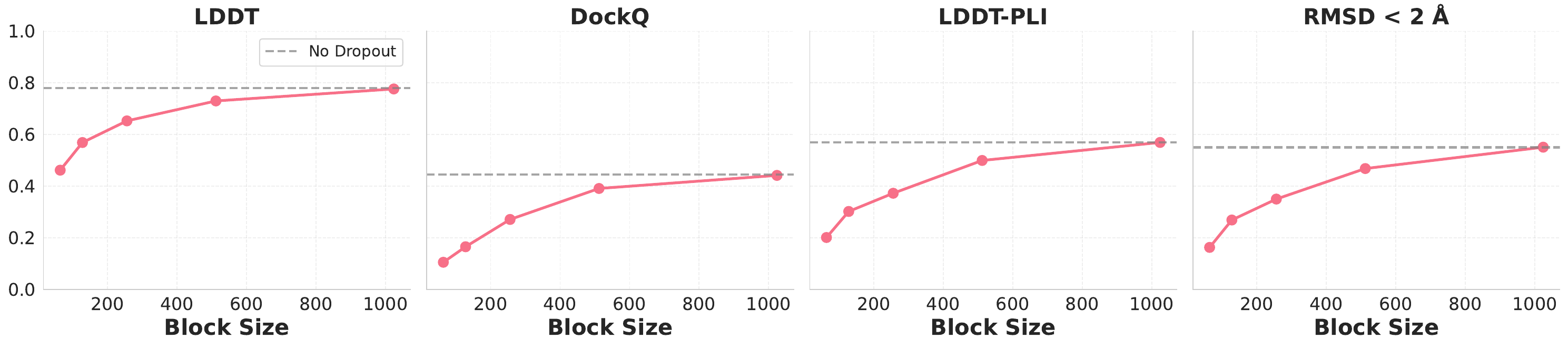}
\caption{
\textbf{\name Performance under blockwise dropout.}
The model is trained with standard triangle multiplication and evaluated under a local blockwise triangle multiplication.
Performance quickly degrades even for local metrics like lDDT.
}
\lblfig{block}
\vspace{-1pt}
\end{figure}

\section{Conclusion}
We introduce \name, a simplified, efficient feature extractor for biomolecular structure prediction.
Models using \name train 1.5$\times$ faster and sample up to $4\times$ faster than those with Pairformer, enabling large-scale, compute-intensive applications of structure prediction.
The key idea is to explicitly materialize a 2-D pair representation, updated via triangle multiplications that capture interactions among residue triplets.
We hypothesize that transforming 1-D sequences into 3-D structures is most effective when mediated through this intermediate pair representation, which naturally encodes distance information.
Triangle multiplication provides a simple and efficient mechanism to do so.

\section{Acknowledgment}
This work was in part supported by the NSF AI Institute for Foundations of Machine Learning (IFML) and UT-Austin Center for Generative AI.
The authors thank Yue Zhao, Sergey Ovchinnikov, Chengyue Gong and Luca Naef for their thoughtful feedback on this manuscript.

\newpage

\bibliography{iclr2026_conference}
\bibliographystyle{iclr2026_conference}

\newpage

\appendix

\section{Architectural Baselines}
\begin{figure}[h]
\centering
\vspace{-5pt}
\begin{subfigure}[t]{0.95\linewidth}
    \centering
    \includegraphics[width=\linewidth]{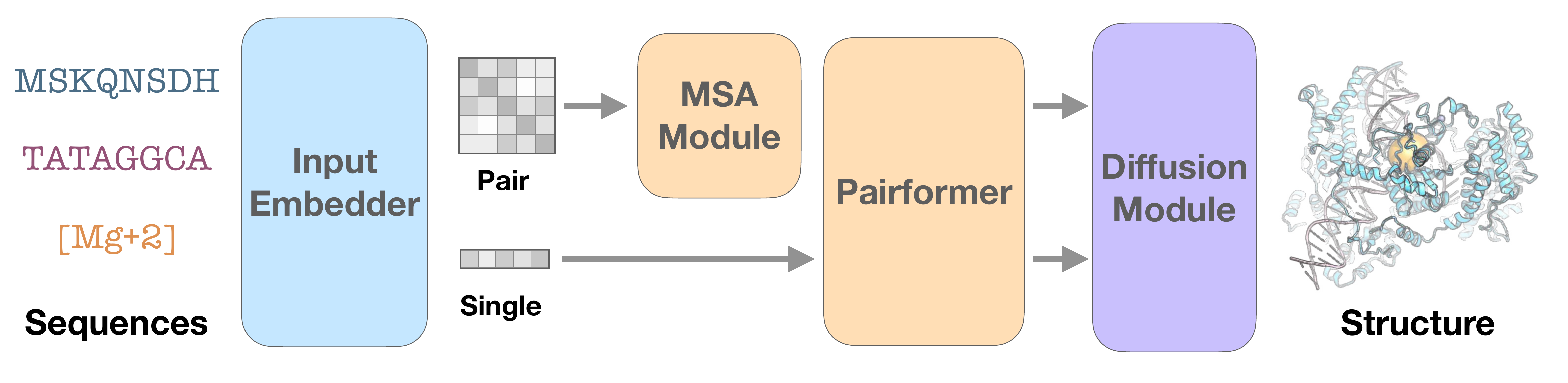}
    \caption{Pairformer-based predictor}
    \label{fig:pipeline_pairformer_v1}
\end{subfigure}
\vspace{6pt}
\begin{subfigure}[t]{0.95\linewidth}
    \centering
    \includegraphics[width=\linewidth]{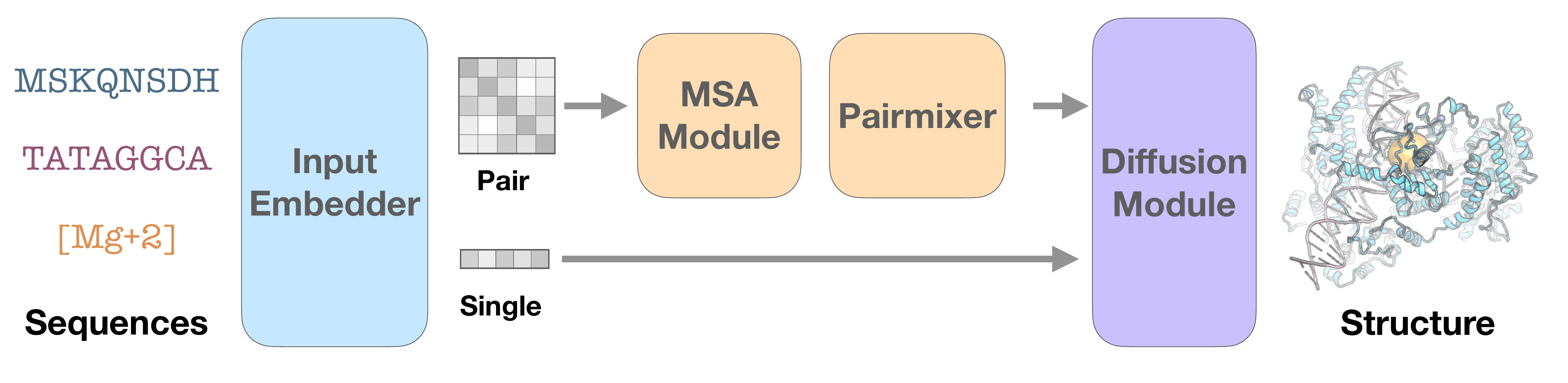}
    \caption{\name-based predictor}
    \label{fig:pipeline_pairmixer_v4}
\end{subfigure}
\vspace{6pt}
\begin{subfigure}[t]{0.95\linewidth}
    \centering
    \includegraphics[width=\linewidth]{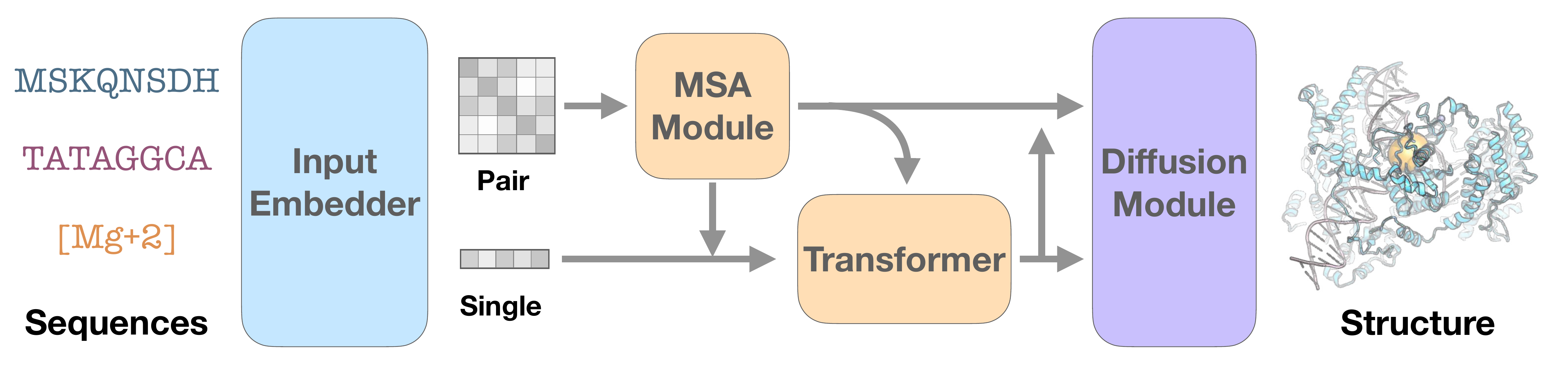}
    \caption{Transformer-based predictor}
    \label{fig:pipeline_transformer}
\end{subfigure}
\vspace{-5pt}
\caption{
\textbf{Overview of biomolecular structure predictors.}
We study the effect of varying backbone architectures while keeping all other modules fixed, except in the Transformer model, where we adjust the connections between the MSA module outputs and the Diffusion module inputs.
}
\label{fig:all_pipelines}
\vspace{-3pt}
\end{figure}

The full cofolding pipeline for all methods can be found at~\reffig{all_pipelines}.

\begin{figure}[t]
\centering
\includegraphics[width=0.99\linewidth]{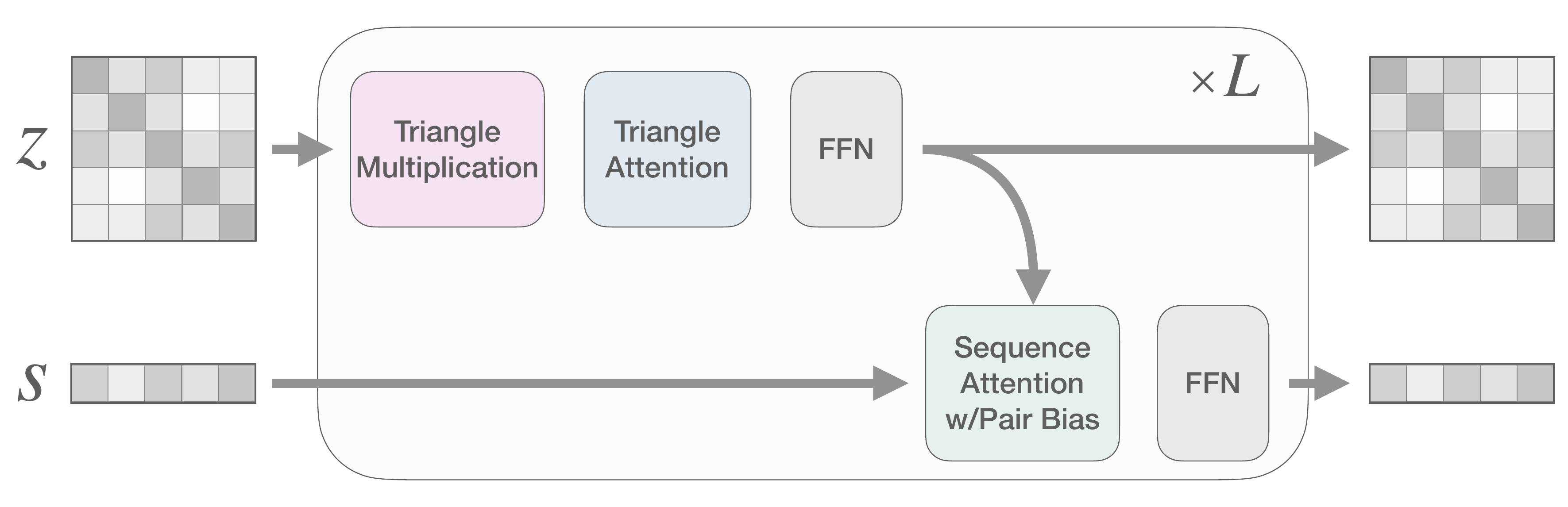}
\begin{subfigure}[t]{0.33\textwidth}
    \centering
    \includegraphics[width=\linewidth]{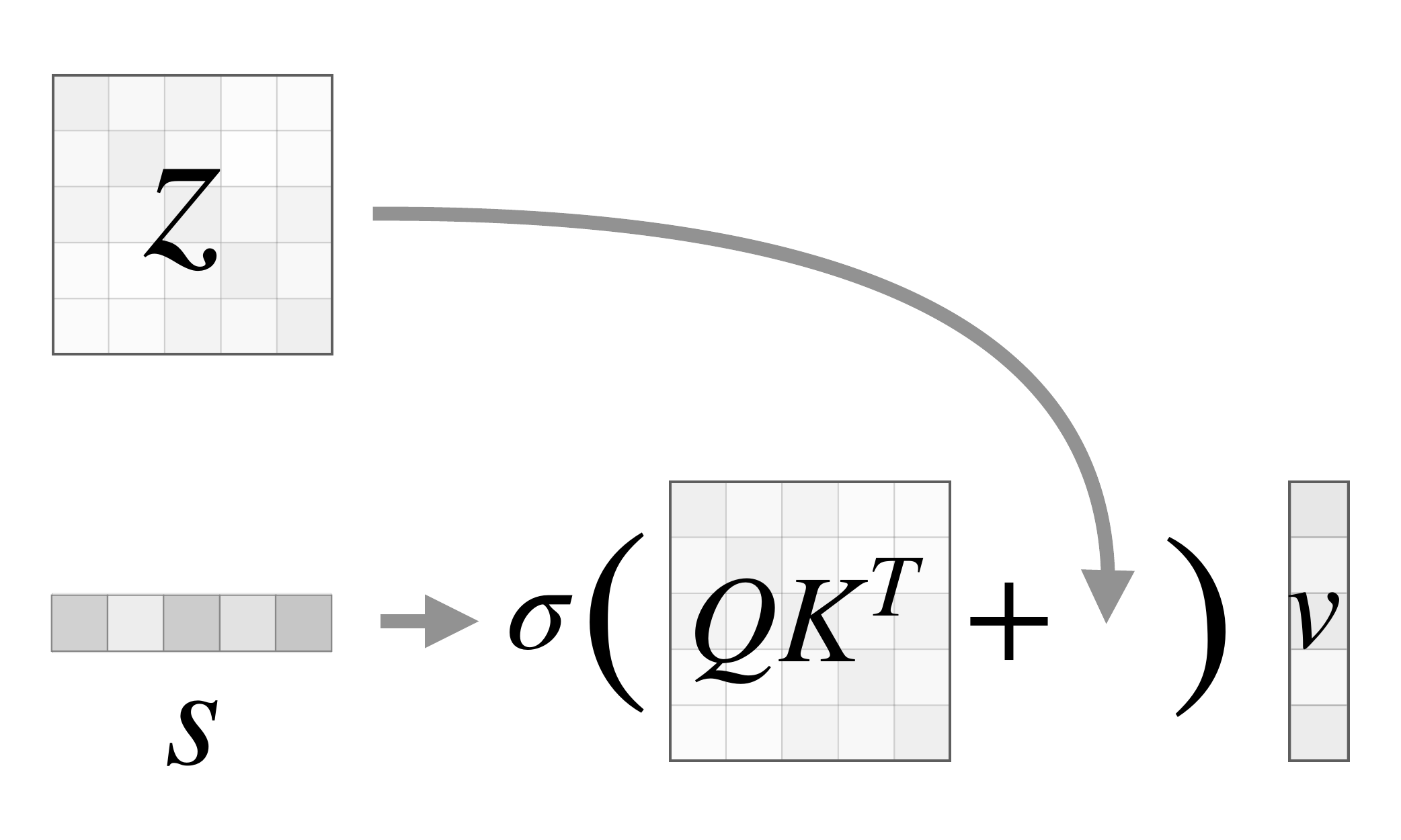}
    \caption{Seq. Attention w/Pair Bias}
    \lblfig{sequence_attention_arch}
\end{subfigure}
\hfill
\begin{subfigure}[t]{0.33\textwidth}
    \centering
    \includegraphics[width=\linewidth]{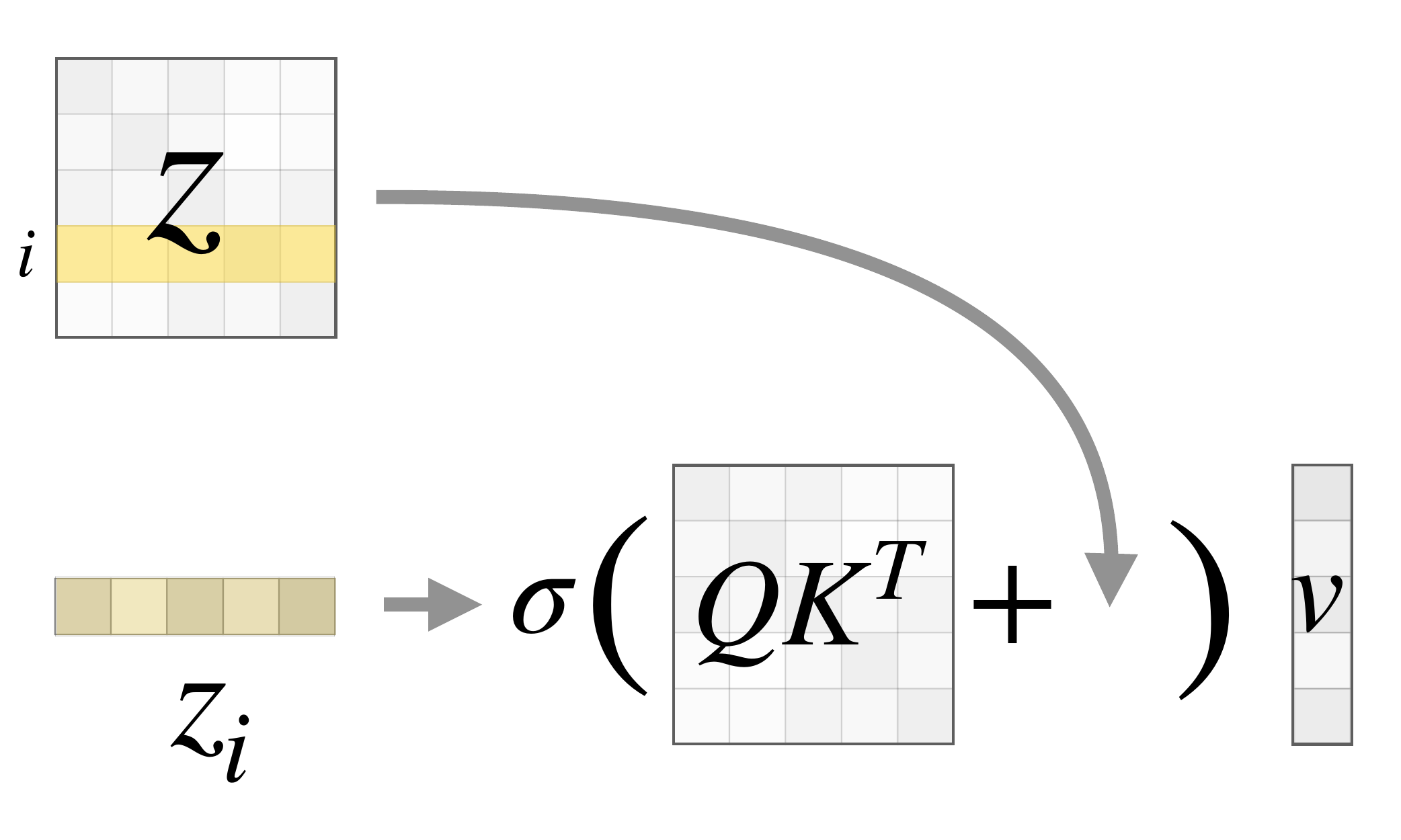}
    \caption{Triangle Attention}    \lblfig{triangle_attention_arch}
\end{subfigure}
\hfill
\begin{subfigure}[t]{0.3\textwidth}
    \centering
    \includegraphics[width=\linewidth]{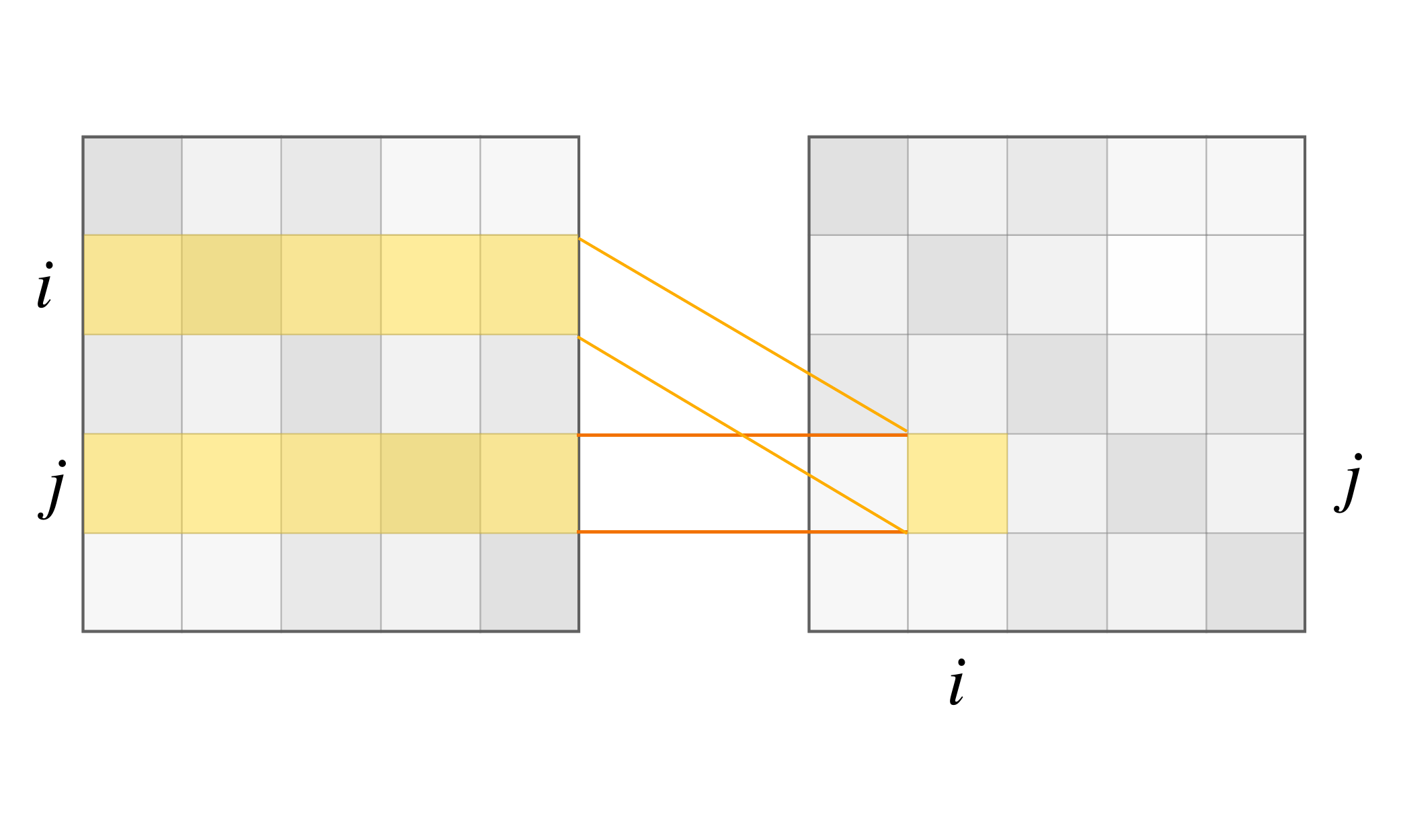}
    \caption{Triangle Multiplication}
    \label{fig:triangle_multiplication_arch}
\end{subfigure}
\caption{
\textbf{Pairformer Architecture and Module Details.}
The main architecture (top) outlines the general Pairformer layer. 
The detailed module architectures (bottom) illustrate the key components: 
(a) Sequence Attention with Pair Bias, (b) Triangle Attention, and (c) Triangle Multiplication modules.
}
\label{fig:pairformer_combined}
\vspace{-1pt}
\end{figure}

\subsection{Pairformer Baseline}
Here we describe the Pairformer architecture of~\reffig{pairformer_combined} in detail.

\paragraph{Attention Primitive.}
The Pairformer extends the standard attention mechanism by incorporating a pairwise bias term derived from the pair representation $\vz$.
Formally, this update is
$$
\begin{aligned}
\mathbf{AttnWithPairBias}(\vx, \vz) 
&= \text{softmax} \Big( (\mW_Q \vx)(\mW_K \vx)^\top + \mW_B \vz \Big)\, \mW_V \vx, \\
\end{aligned}
$$
where $\vx \in \mathbb{R}^{L \times C_x}$ is a sequence representation, $\vz \in \mathbb{R}^{L \times L \times C_z}$ is a pair representation, $(\mW_Q, \mW_K, \mW_V)$ are standard attention projection matrices, and $\mW_B$ projects the pair representation into an attention bias term~\footnote{single head and removed scaling for brevity}.

\paragraph{The Sequence Update} first performs attention with pair bias (see~\reffig{sequence_attention_arch}) and then applies a feed-forward network.
At layer $l$, we compute the update
$$
\begin{aligned}
\tilde{\vs}_{l+1} &= \vs_l + \mathbf{AttnWithPairBias}(\vs_l, \vz_l) \\
\vs_{l+1} &= \tilde{\vs}_{l+1} + \mathbf{FFN}(\tilde{\vs}_{l+1})
\end{aligned}
$$

\paragraph{The Pair Update} mixes the tokens in pair representation $\vz \in \mathbb{R}^{L \times L \times C_z}$ using triangle attention and triangle multiplication, then applies a feedforward network.

The \textit{Triangle Attention} operates on each row of the pair representation $\vz_i \in \mathbb{R}^{L \times C_z}$ as an independent sequence, applying sequence attention with pair bias to each row separately\footnote{In practice, another layer of triangle attention is performed on the columns.} (see~\reffig{triangle_attention_arch}).
Formally, the update for row $i$ is defined as
$$\mathbf{TriAttn}(\vz)_i = \mathbf{AttnWithPairBias}(\vz_i, \vz) $$

The \textit{Triangle Multiplication} integrates features across different rows of the pair representation~\footnote{In practice, another layer of triangle multiplication is performed on the columns.} (see~\reffig{triangle_multiplication_arch}).
Formally, the update for feature $\vz_{ij}$ is defined as 
$$\mathbf{TriMul}(\vz)_{ij} = \sum_{k=1}^L ( \mW_a \vz_{ik}) \odot (\mW_b \vz_{jk})$$
where $\mW_a, \mW_b$ are linear projection layers. 

Both pair operations were introduced to reason over triplets of residues, intuitively enabling the model to learn to follow geometric constraints in 3-D space~\citep{jumper2021alphafold2}.

\subsection{Transformer Baseline}
\lblsec{transformer_baseline}
Our transformer baseline removes the pair update from the Pairformer and keeps only the sequence update.
We also modify the MSA module to make it more effective with the transformer baseline.
Instead of outputting only $\vz^{\text{msa}}$, it produces an additional sequence representation $\vs^{\text{msa}}$, obtained by indexing the first row of the processed MSA representation.
This $\vs^{\text{msa}}$ is fed into the transformer, while $\vz^{\text{msa}}$ serves as the pair bias.
Additionally, the diffusion module expects both sequence and pair representations.
Because the pair features are otherwise less processed in this baseline, we update them with the outer sum of the sequence representation.
Formally,
$$
\vz^{\text{backbone}}_{ij} =  \vz^{\text{msa}}_{ij} + \mW_{\vs \rightarrow \vz} \vs^{\text{backbone}}_i + \mW_{\vs \rightarrow \vz} \vs^{\text{backbone}}_j 
$$
where $\mW_{\vs \rightarrow \vz} \in \mathbb{R}^{C_z \times C_s}$ is a projection layer.
This is illustrated in~\reffig{pipeline_transformer}.

\section{FLOPs Calculations}
\lblsec{appendix/flops}

\begin{table}[t]
\centering
\small
\begin{tabular}{@{}l r@{}}
\toprule
\textbf{Module / Operation} & \textbf{FLOPs} \\
\midrule
\textbf{Backbone / MSA Module} & \\
\quad \textbf{Pair Update} & \\
\quad\quad \textbf{Triangle Attention} & \\
\quad\quad\quad Matrix Multiply & $8\,L^{3}C_z$ \\
\quad\quad\quad Projection & $20\,L^2C_z^{2}$ \\
\quad\quad \textbf{Triangle Multiplication} & \\
\quad\quad\quad EinSum & $4\,L^{3}C_z$ \\
\quad\quad\quad Projection & $24\,L^2 C_z^{2}$ \\
\quad\quad \textbf{Pair FFN} & $24\,L^2 C_z^{2}$ \\[3pt]
\quad \textbf{Sequence Update} & \\
\quad\quad \textbf{Sequence Attention (with Pair Bias)} & \\
\quad\quad\quad Matrix Multiply & $4\,L^2C_s$ \\
\quad\quad\quad Projection & $10\,LC_s^{2}$ \\
\quad\quad \textbf{Sequence FFN} & $24\,LC_s^{2}$ \\
\midrule
\textbf{Diffusion Transformer} & \\
\quad Attention (Pair Bias) – Matrix Multiply & $4\,L^2C_a$ \\
\quad Attention (Pair Bias) – Projection & $10\,LC_a^{2}$ \\
\quad Sequence FFN & $16\,LC_a^{2}$ \\
\midrule
\textbf{Full Modules} & \\
\quad MSA Module & $R \,D_m \,(12 L^3 C_z + 68 L^2 C_z^2)$ \\
\quad Pairformer & $R \,D_p \,(12 L^3 C_z + 68 L^2 C_z^2 + 4 L^2 C_s + 34 L C_s^2)$ \\
\quad Structure Module & $M \,D_d \,(4 L^2 C_a + 26 L C_a^2)$ \\
\bottomrule
\end{tabular}
\caption{
\textbf{Breakdown of FLOPs in \af architectural components.}
Variables: $L=\texttt{max\_tokens}$, $C_z=\texttt{token\_z}$, 
$C_s=\texttt{token\_s}$, $C_a=2\times\texttt{token\_z}$, 
$R=\texttt{recycles}$, $D_p=\texttt{pairformer\_depth}$, 
$D_m=\texttt{msa\_depth}$, $D_d=\texttt{diffusion\_depth}$, 
$M=\texttt{multiplicity}$.
}
\lbltbl{flops}
\end{table}

Our biomolecular structure predictor uses a multi-resolution transformer that denoises atom coordinates at both the token and heavy-atom levels (see~\reffig{pipeline}).
In this design, a backbone refines token representations, which are then processed by a conditional diffusion transformer. The backbone runs once per sequence, while the diffusion transformer can generate arbitrarily many samples.

In~\reftbl{flops}, we present the mathematical FLOP calculations for each component, and in~\reftbl{boltz1_hyperparameters} we report the total training and inference FLOPs for all model architectures.

\paragraph{Boltz-1 Hyperparameters}
The Boltz-1 architecture is defined by several key components and hyperparameters that influence its performance.
We identify the following set of critical hyperparameters:
\begin{itemize}
    \item \textbf{Input}: The input is defined by the number of input tokens ($L$), the single token dimension ($C_s$), and the pair token dimension ($C_z$).
    \item \textbf{Feature extractor}: The feature extractor consists of Pairformer and MSA blocks that process single and pair representations; its configuration is determined by the number of Pairformer blocks $D_p$, MSA blocks $D_m$.
    \item \textbf{Diffusion model}: The diffusion model is a transformer architecture made up of Multi-Head Attention (MHA) transformer layers. Its configuration is determined by the number of diffusion blocks ($D_d$) and the widths of its layers $C_a = 2\, C_z$.
\end{itemize}

\paragraph{Feature extractors.}
The feature extractors is a concatenation of $D_m$ MSA blocks and $D_p$ pairformer blocks.
Each pairformer block primarily consists of two parallel update paths: the pair representation path and the single representation path (see~\reffig{pairformer_combined}).
Each path is further processed by a FFN. The pair representation path includes two \textit{triangular self-attention} updates and two \textit{triangular multiplication} updates (applied row-wise and column-wise). These are analogous to axial attention mechanisms~\citep{ho2019axial} operating over an $L \times L$ pair matrix, where each attention pass involves computations along one length-$L$ dimension for each of the $L$ rows or columns.

Each pair of triangular attention pass incurs a computational cost of $O(8 L^3 C_z)$ FLOPs. 
The triangle multiplication einsum operations require a quadratic FLOPs term per input token (total FLOPs of $4 L^3 C_z$).
Following the triangle updates, a feed-forward network (FFN) is applied to each pair representation entry. The single representation path also contributes to the computational load, but its cost is quadratic in $L$.

Each MSA block is lighter than the full pairformer blocks and consists of a pair of triangular attention layers and a pair of triangular operations, followed by a FFN network for pair representation FFN ($C_z$), but without a single representation FFN and attention with pair bias.
It also includes an additional OuterProductMean and pair-weighted averaging on the MSA, which we omit from our FLOPs calculations.

\paragraph{Diffusion Model.}
Each diffusion module block resembles a standard transformer block with a standard \textit{self-attention} mechanism and a conditioning block. As with the trunk block analysis, we ignore bias terms, gating, and layer normalization for simplicity. We also ignore the cost of Atom Attention Encoder and Atom Attention Decoder that run on atoms, since those modules adopt sequence-local attention ~\citep{wohlwend2024boltz1} and their computational cost is negligible.
The conditioned transition block of the diffusion model is dominated by dense matrix multiplications that scale quadratically with the hidden size $C_a$. The bulk of the compute arises from the SwiGLU feed-forward pathway, which contributes both a pair of linear projections
($4\,C_a^{2}$) and the associated activation matmul ($2\,C_a^{2}$). In addition, cross--path transformations are introduced via the $a \!\to\! b$ and $b \!\to\! a$ projections (each $2\,C_a^{2}$), followed by an output projection ($2\,C_a^{2}$). Finally, the gating mechanisms for both the $a$ and $b$ streams contribute
another $2\,C_a^{2}$ apiece. The total FLOPs per structure block can therefore be approximated as the sum of the attention, MatMuls, and feed-forward components (see~\reftbl{flops}).

\section{Additional Results}

\subsection{System-level comparisons on the CASP15 dataset}
\lblsec{casp_results}

We report results on the CASP15 dataset in~\reftbl{casp15_against_literature}.
These numbers differ slightly from~\reftbl{full_results_casp15} because we further filter proteins to ensure all methods are evaluated on the same set.

\begin{figure}[h]
\centering
\includegraphics[width=0.99\linewidth]{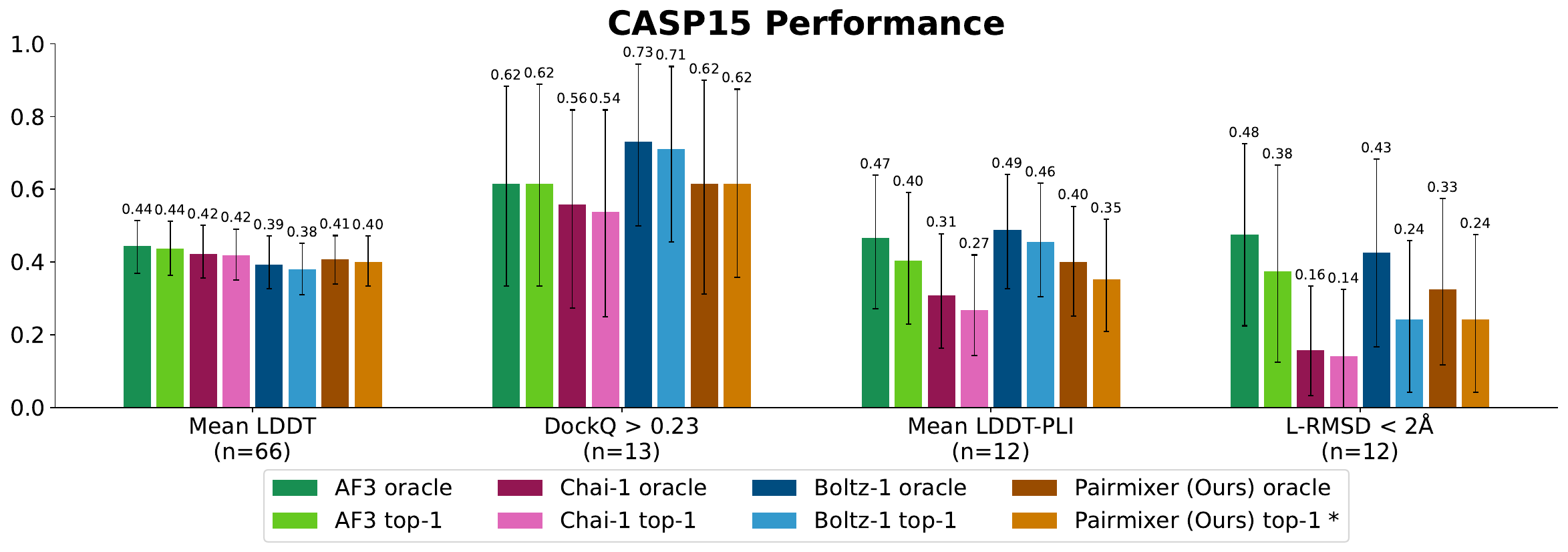}
\caption{
\textbf{System-level comparison on the CASP15 test set.}
We evaluate against AlphaFold3, Chai-1, and Boltz-1 on protein and small-molecule structure prediction. 
\textcolor{brown}{\name} performs competitively with these state-of-the-art approaches. 
Error bars denote bootstrapped 95\% confidence intervals.
$^*$Since we do not train a confidence model, results are reported using the first prediction. 
}
\lbltbl{casp15_against_literature}

\end{figure}

\clearpage

\subsection{Full biomolecular structure prediction results for RCSB and CASP15}
\reftbl{full_results_rcsb} and~\reftbl{full_results_casp15} report the full set of evaluation metrics across all architectures, along with the number of complexes evaluated by each metric.
We retrain Boltz-1 for our Pairformer baselines and additionally include comparisons against the public checkpoint.

\begin{table}[h]
\centering
\caption{
\textbf{Model Performance on the Boltz RCSB test set. }
The metric is computed on the best-performing protein out of five samples (oracle).
}
\tiny
\begin{tabular}{lcccccccc}
Architecture & Epoch & GPU-Days & lDDT & $\text{DOCKQ}_{>0.23}$ & $\text{DOCKQ}_{>0.49}$ & $\text{lDDT}_\text{PLI}$ & $\text{RMSD}_{<1}$ & $\text{RMSD}_{<2}$ \\
 & & & (n=539) & (n=342) & (n=342) & (n=250) & (n=250) & (n=250) \\
\shline
\multicolumn{9}{l}{\textbf{Small }} \\
Transformer & 68 & 86 & 0.68 & 0.51 & 0.35 & 0.47 & 0.32 & 0.43 \\
Pairformer (Boltz-1) & 68 & 125 & 0.74 & 0.58 & 0.44 & 0.52 & 0.37 & 0.48 \\
Pairmixer (Ours) & 68 & 98 & 0.73 & 0.59 & 0.44 & 0.51 & 0.33 & 0.45 \\
\midrule
\multicolumn{9}{l}{\textbf{Medium }} \\
Transformer & 68 & 128 & 0.67 & 0.50 & 0.36 & 0.47 & 0.33 & 0.46 \\
Pairformer (Boltz-1) & 68 & 194 & 0.75 & 0.60 & 0.47 & 0.53 & 0.36 & 0.49 \\
Pairmixer (Ours) & 68 & 146 & 0.76 & 0.60 & 0.46 & 0.54 & 0.40 & 0.53 \\
\midrule
\multicolumn{9}{l}{\textbf{Large }} \\
Transformer & 68 & 173 & 0.69 & 0.51 & 0.37 & 0.48 & 0.33 & 0.46 \\
Pairformer (Boltz-1) & 68 & 290 & 0.76 & 0.61 & 0.49 & 0.54 & 0.41 & 0.52 \\
Pairmixer (Ours) & 68 & 192 & 0.75 & 0.61 & 0.46 & 0.55 & 0.38 & 0.51 \\
\midrule
\multicolumn{9}{l}{\textbf{Large Phase 2}} \\
Transformer & 20 & 232 & 0.70 & 0.53 & 0.38 & 0.51 & 0.35 & 0.48 \\
Pairformer (Boltz-1) & 20 & 421 & 0.78 & 0.64 & 0.50 & 0.57 & 0.44 & 0.54 \\
Pairmixer (Ours) & 20 & 269 & 0.78 & 0.63 & 0.49 & 0.57 & 0.45 & 0.55 \\
\midrule
\multicolumn{9}{l}{\textbf{Boltz-1 public model}} \\
Pairformer (Boltz-1) & - & - & 0.79 & 0.64 & 0.51 & 0.58 & 0.46 & 0.57 \\
\end{tabular}
\lbltbl{full_results_rcsb}
\end{table}

\begin{table}[h]
\centering
\caption{
\textbf{Model Performance on CASP15 test set. }
The metric is computed on the best-performing protein out of five samples (oracle).
}
\tiny
\begin{tabular}{lcccccccc}
Architecture & Epoch & GPU-Days & lDDT & $\text{DOCKQ}_{>0.23}$ & $\text{DOCKQ}_{>0.49}$ & $\text{lDDT}_\text{PLI}$ & $\text{RMSD}_{<1}$ & $\text{RMSD}_{<2}$ \\
 & & & (n=66) & (n=14) & (n=14) & (n=12) & (n=12) & (n=12) \\
\shline
\multicolumn{9}{l}{\textbf{Small }} \\
Transformer & 68 & 86 & 0.35 & 0.22 & 0.17 & 0.21 & 0.06 & 0.10 \\
Pairformer (Boltz-1) & 68 & 125 & 0.39 & 0.46 & 0.24 & 0.36 & 0.10 & 0.21 \\
Pairmixer (Ours) & 68 & 98 & 0.37 & 0.39 & 0.21 & 0.35 & 0.06 & 0.16 \\
\midrule
\multicolumn{9}{l}{\textbf{Medium }} \\
Transformer & 68 & 128 & 0.35 & 0.19 & 0.16 & 0.27 & 0.04 & 0.15 \\
Pairformer (Boltz-1) & 68 & 194 & 0.38 & 0.66 & 0.35 & 0.39 & 0.14 & 0.23 \\
Pairmixer (Ours) & 68 & 146 & 0.39 & 0.49 & 0.39 & 0.38 & 0.12 & 0.24 \\
\midrule
\multicolumn{9}{l}{\textbf{Large }} \\
Transformer & 68 & 173 & 0.36 & 0.29 & 0.16 & 0.26 & 0.06 & 0.10 \\
Pairformer (Boltz-1) & 68 & 290 & 0.41 & 0.68 & 0.43 & 0.37 & 0.12 & 0.31 \\
Pairmixer (Ours) & 68 & 192 & 0.38 & 0.50 & 0.35 & 0.34 & 0.12 & 0.23 \\
\midrule
\multicolumn{9}{l}{\textbf{Large Phase 2}} \\
Transformer & 20 & 232 & 0.37 & 0.34 & 0.17 & 0.26 & 0.11 & 0.11 \\
Pairformer (Boltz-1) & 20 & 421 & 0.42 & 0.64 & 0.43 & 0.36 & 0.10 & 0.28 \\
Pairmixer (Ours) & 20 & 269 & 0.41 & 0.52 & 0.36 & 0.34 & 0.14 & 0.31 \\
\midrule
\multicolumn{9}{l}{\textbf{Boltz-1 public model}} \\
Pairformer (Boltz-1) & - & - & 0.4 & 0.68 & 0.43 & 0.45 & 0.23 & 0.42 \\
\end{tabular}
\lbltbl{full_results_casp15}
\end{table}

\subsection{Details for Diverse Biomolecular Structure Prediction}
\lblsec{diverse_task_details}
We evaluate \name across several benchmark datasets listed in~\reftbl{more_benchmarks}. 
Below, we describe the dataset preparation and evaluation protocols used for these benchmarks.

\textbf{Protein-ligand complexes.}
We evaluate performance on protein–ligand complexes using the PoseBuster benchmark.
The original dataset contains 428 complexes.
Applying a training-date cutoff of September 30, 2021 reduces this to 373, and after removing redundant protein–ligand complexes, the final benchmark includes 298 structures.
Evaluation uses standard protein–ligand metrics, including RMSD $<$ 2Å, RMSD $<$ 1Å, and protein–ligand lDDT.
We compare a Pairmixer model finetuned for longer to the publicly available Boltz-1 checkpoint with Pairformer in~\reftbl{posebusters}.
Under this setup, \name performs comparably to Pairformer, with at most a 1\% drop in performance, while the architecture is significantly simpler and more efficient, requiring no attention in the backbone.

\textbf{Antibody–antigen complexes.}
We evaluate on the antibody–antigen benchmark introduced in the AlphaFold3 paper.
First, we extract the relevant chains from the publicly released AlphaFold3 files and remove unresolved residues.
Of the 71 total complexes, 70 pass our data pipeline.
The protein-protein interface is evaluated using the DockQ $>$ 0.23 metric.
We compare large Pairformer, Pairmixer, and Transformer models trained for the same number of iterations in~\reftbl{antibodies}.
We find that \name matches the performance of Pairformer (0.23), while the sequence-only Transformer performs substantially worse (0.08).

\textbf{Protein–nucleic acid and RNA-only complexes.}
We evaluate our models on the protein–nucleic acid dataset from the AlphaFold-3 paper.
Of the 199 structures, 172 pass our data pipeline.
We also consider a subset of 27 RNA-only structures.
For RNA-only complexes, we assess folding quality using lDDT, while for protein–nucleic acid complexes we evaluate interface accuracy using Interface Contact Similarity (ICS) and Interface Patch Similarity (IPS).
We compare large Pairformer, Pairmixer, and Transformer models trained for the same number of iterations in~\reftbl{rna_only} and~\reftbl{protein_nucleic_acids}.
On protein–nucleic acid complexes, \name performs comparably to Pairformer, with the Transformer lagging behind.
For RNA-only structures, \name again matches Pairformer, while the Transformer performs better, likely due to limited RNA structural training data.
Notably, Pairmixer achieves performance comparable to Pairformer despite removing sequence attention from the trunk.

\begin{figure}[t]
\centering
\begin{subfigure}[b]{0.4\linewidth}
    \centering
    \includegraphics[width=\linewidth]{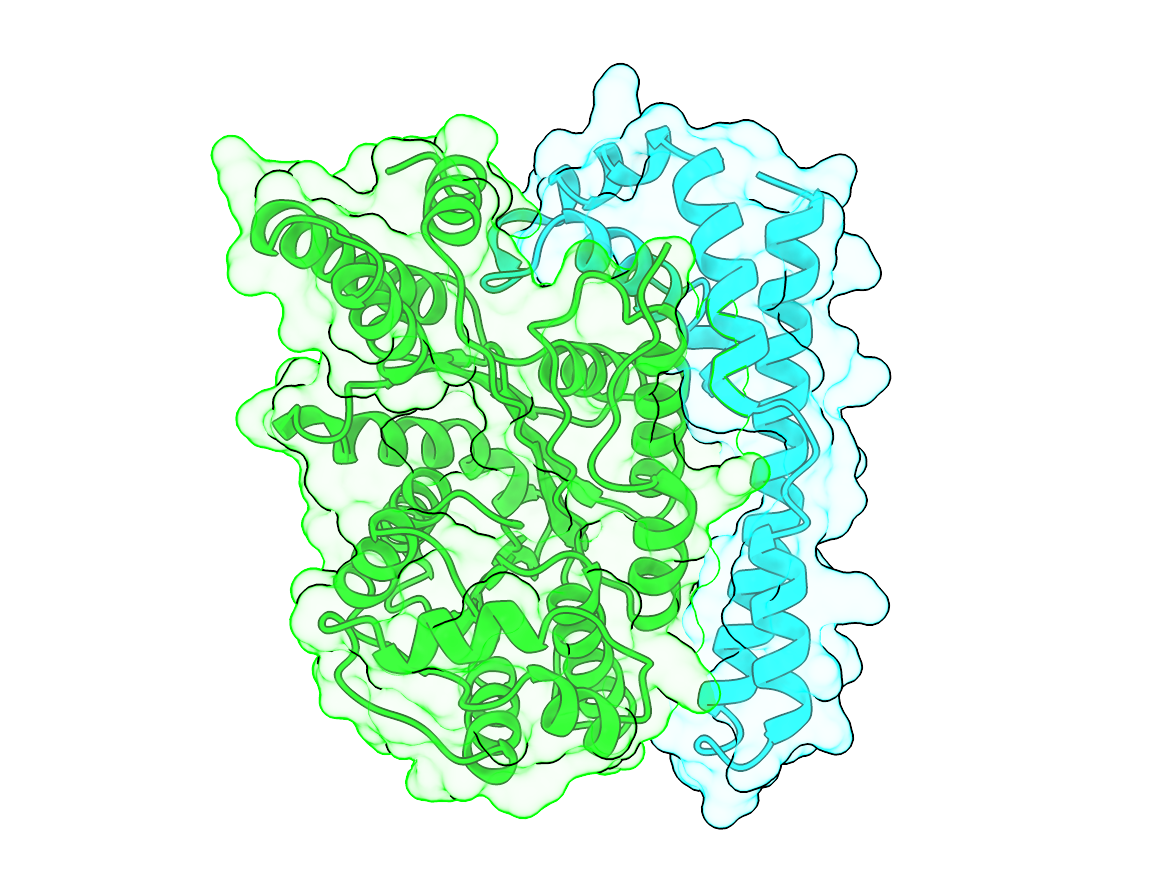}
    \caption{Pairformer-based predictions}
\end{subfigure}
\hspace{0.05\linewidth} %
\begin{subfigure}[b]{0.4\linewidth}
    \centering
    \includegraphics[width=\linewidth]{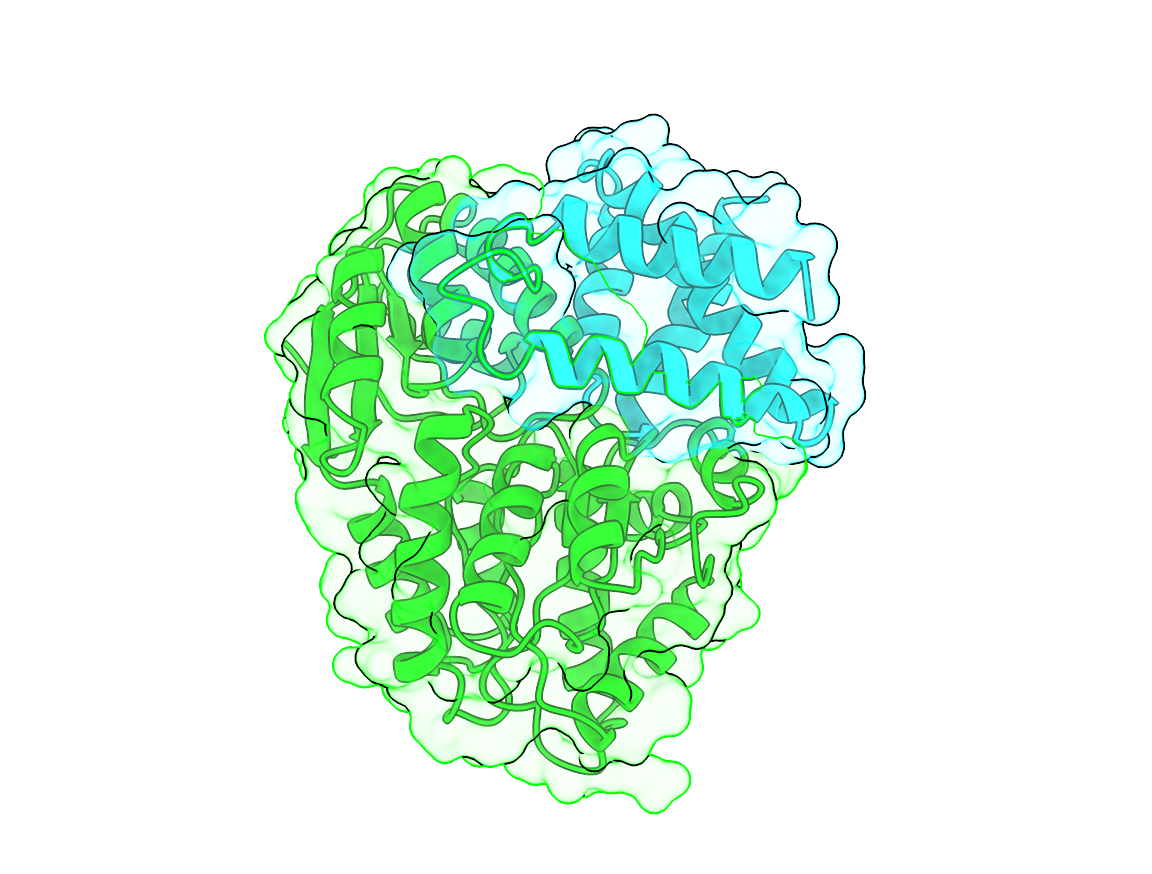}
    \caption{\name-based predictions}
\end{subfigure}
\caption{
\textbf{Qualitative visualizations of de-novo binders.}
Target is shown in green and binder is shown in blue.
PDB code: 1P5J
}
\vspace{-1em}
\label{fig:bindfast_qual}
\end{figure}

\subsection{Details of Pairmixer applied to Protein Design}
\lblsec{bindfast_details}

BindCraft~\citep{pacesa2025bindcraft}, BoltzDesign~\citep{cho2025boltzdesign1}, and hallucination-based protein design methods~\citep{frank2024scalable,wicky2022hallucinating,jendrusch2025alphadesign,goverde2023novo,bryant2022evobind,anishchenko2021novo} have recently demonstrated that structure predictors can be repurposed as differentiable scoring functions for sequence optimization.
The input sequence is treated as a set of learnable parameters and is updated by backpropagating through a structure predictor, thereby jointly refining sequence and structure toward favorable interactions with the target protein or small molecule. 
While powerful, these methods have practical limitations: memory demands are high and sequence generation is slow, requiring hundreds of runs of the structure predictor per design.
This inefficiency makes the approach prohibitively expensive, particularly for larger systems.

To address these challenges, we introduce BindFast, a scalable and efficient framework for binder design which replaces BoltzDesign’s Pairformer backbone with \name.
BindFast substantially reduces the runtime and memory footprint of binder generation and aim to accelerate the discovery of high-quality binders, particularly for large targets.

In~\reftbl{bindfast}, we benchmark the runtime performance of BindFast against BoltzDesign for generating 110-residue binders across a range of target proteins with biotechnological relevance, using an A100 GPU with 80 GB memory. 
BoltzDesign failed with out-of-memory (OOM) errors on targets larger than 500 residues, whereas BindFast extended this limit to 650 residues, a 30\% improvement in target size. 
For protein targets where both models executed without memory overflow, BindFast achieves speedups of 2x to 2.6x at total sequence lengths ranging from 140 to 607, respectively. 
Qualitative comparisons in~\reffig{bindfast_qual} further indicate that BindFast produces designs comparable to those of BoltzDesign, underscoring its potential for faster in-silico iteration and enabling the design of binders against larger, more biologically relevant targets.

\section{Additional Analysis}

\subsection{Pairformer Ablations}

We performed ablation experiments on a small 12-layer Pairformer model to isolate the contributions of triangle multiplication, triangle attention, and sequence updates in~\reftbl{pairformer_ablation}.
The results show that, under a short training schedule of 60 epochs (3M samples), both triangle multiplication and triangle attention are essential for performance, while sequence updates have minimal impact.
Notably, \name recovers performance with additional training.

\begin{table}[t]
\centering
\caption{
\textbf{Pairformer Ablation.}
We remove each module in the Pairformer one at a time.
}
\begin{tabular}{llccccc}
Ablation & GPU days & lDDT & $\text{DOCKQ}_{>0.23}$ & $\text{lDDT}_\text{PLI}$ & $\text{RMSD}_{<2}$ \\
\shline
-  & 82 & 0.74 & 0.57 & 0.52 & 0.50 \\
No Seq Update & 80 & 0.73 & 0.57 & 0.54 & 0.49 \\
No Tri Att & 66 & 0.70 & 0.55 & 0.50 & 0.48 \\
No Tri Mul & 71 & 0.70 & 0.53 & 0.49 & 0.46 \\
\end{tabular}
\lbltbl{pairformer_ablation}
\end{table}

\begin{table*}[b]
\vspace{-.5em}
\centering
\caption{
\textbf{\name ablations experiments.}
Default settings are marked in grey.
See~\refsec{additional_ablations} for details.
$D_p$: number of pairmixer layers.
$D_d$: number of diffusion transformer layers.
}
\begin{subtable}[t]{0.49\linewidth}
\caption{\textbf{FFN Hidden Dimension}}
\vspace{-5pt}
\centering
\tablestyle{5pt}{1.05}
\begin{tabular}{p{20pt} p{22pt} p{32pt} p{24pt} p{24pt}}
dim & \tiny{lDDT} & \tiny{$\text{DOCKQ}_{>0.49}$} & \tiny{$\text{lDDT}_\text{PLI}$} & \tiny{$\text{RMSD}_{<1}$} \\
\shline
256  &  0.71 & 0.38 & 0.50 & 0.34 \\
512  & \baseline{0.71} & \baseline{\textbf{0.42}} & \baseline{0.50} & \baseline{0.33} \\
1024 &  \textbf{0.74} & 0.40 & \textbf{0.53} & \textbf{0.35} \\
\end{tabular}
\lbltbl{ffndim}
\end{subtable} %
\begin{subtable}[t]{0.49\linewidth}
\caption{\textbf{Triangle Mul Dimension}}
\vspace{-5pt}
\centering
\tablestyle{5pt}{1.05}
\begin{tabular}{p{20pt} p{22pt} p{32pt} p{24pt} p{24pt}}
dim & \tiny{lDDT} & \tiny{$\text{DOCKQ}_{>0.49}$} & \tiny{$\text{lDDT}_\text{PLI}$} & \tiny{$\text{RMSD}_{<1}$} \\
\shline
64  & 0.71 & 0.41 & 0.50 & 0.34 \\
128  & \baseline{0.71} & \baseline{\textbf{0.42}} & \baseline{0.50} & \baseline{0.33} \\
256 & \textbf{0.73} & \textbf{0.42} & \textbf{0.52} & \textbf{0.37} \\
\end{tabular}
\lbltbl{trimuldim}
\end{subtable} %

\vspace{5pt}  %

\begin{subtable}[t]{0.5\linewidth}
\caption{\textbf{Mixing Method}}
\vspace{-5pt}
\lbltbl{mixer}
\centering
\tablestyle{5pt}{1.05}
\begin{tabular}{p{50pt} p{20pt} p{32pt} p{20pt} p{20pt}}
mixer & \tiny{lDDT} & \tiny{$\text{DOCKQ}_{>0.49}$} & \tiny{$\text{lDDT}_\text{PLI}$} & \tiny{$\text{RMSD}_{<1}$} \\
\shline
FFT & 0.66 & 0.34 & 0.45 & 0.27 \\
AvgPool & 0.69 & 0.35 & 0.48 & 0.31 \\
TriMul-rows & 0.70 & 0.35 & 0.49 & 0.32 \\
TriMul-both  & \baseline{\textbf{0.71}} & \baseline{\textbf{0.42}} & \baseline{\textbf{0.50}} & \baseline{\textbf{0.33}} \\
\end{tabular}
\end{subtable} %
\begin{subtable}[t]{0.45\linewidth}
\caption{\textbf{Diffusion Transformer Depth}}
\vspace{-5pt}
\lbltbl{other_structure_model}
\centering
\tablestyle{6pt}{1.05}
\begin{tabular}{p{10pt} p{10pt} p{20pt} p{32pt} p{20pt} p{20pt}}
$D_p$ & $D_d$ & \tiny{lDDT} & \tiny{$\text{DOCKQ}_{>0.49}$} & \tiny{$\text{lDDT}_\text{PLI}$} & \tiny{$\text{RMSD}_{<1}$} \\
\shline
12 & 12 & 0.73 & 0.43 & 0.52 & 0.34 \\
24 & 24 & \baseline{\textbf{0.75}} & \baseline{\textbf{0.45}} & \baseline{\textbf{0.54}} & \baseline{\textbf{0.40}} \\  %
\end{tabular}
\end{subtable} %
\lbltbl{ablations}
\end{table*}

\subsection{Additional Ablations}
\lblsec{additional_ablations}
\paragraph{Triangle Multiplication vs. Feed-Forward Network. }
We aim to understand how the performance is affected by the triangle multiplication and pair feed-forward networks, the two core ingredients of the \name architecture.
In~\reftbl{ffndim} and~\reftbl{trimuldim}, we vary the hidden dimensions of these components to evaluate model's sensitivity.
For the FFN, we change the hidden dimension that the model expands to.
For triangle multiplication, we instead project the features into higher- or lower-dimensional spaces before the multiplication and then project them back to the input dimension.
We find that decreasing the FFN hidden dimension does not change performance much, while doubling the FFN dimension increases the mean lDDT from 0.71 to 0.74.
We see a similar trend with triangle multiplication dimensions -- doubling the hidden dimension improves the mean lDDT from 0.71 to 0.73, while reducing the dimensionality does not change lDDT.

\paragraph{Other mixing methods. }
Triangle multiplication mixes features within the $\vz\in \mathbb{R}^{L \times L \times D}$ pair representation.
In~\reftbl{mixer}, we replace this operation with alternative, simpler mixing functions.
First, we ablate the outgoing triangle multiplications, retaining only the incoming variant.
Second, we introduce an FFT mixer that applies the discrete Fourier transform along rows and columns, following FNet~\citep{lee2021fnet}.
Finally, we test a pooling mixer that averages representations across each row (and column) and adds the result back to all positions along the corresponding axis.

We find that these simplified approaches are insufficient and underperform compared to vanilla triangle multiplication.
For instance, the FFT mixer likely fails because it mixes features solely based on sequence position, ignoring discontinuities introduced by multiple chains.

\paragraph{Diffusion Module. }
The diffusion module takes the latent representations as input and decodes the 3-dimensional protein structure using a 24-layer transformer.
In~\reftbl{ablations}, we evaluate how sensitive the Pairformer and \name architectures are to the size of the diffusion module.

\subsection{Triangle Multiplication applied to the Match3 Task}

To separate the effect of quadratic pair representations from the cubic cost of triangle operations, we evaluate these components on the Match3 task, a benchmark for learning 3-way interactions~\citep{sanford2023match3,kozachinskiy2025strassen}.
Match3 gives the model a sequence $\mathbf{x}\in [M]^N$ and asks whether any triple of distinct elements sums to zero mod $M$.
We use $N=16$, $M=64$, a hidden dimension of 8, comparable parameter counts (900–1100), and standard embedding, projection, and max-pooling layers, training on balanced datasets.

We compare three architectures that separate representational and computational factors: standard self-attention (linear representations, quadratic compute), third-order self-attention (linear representations, cubic compute~\citep{roy2025simplicial}), and triangle multiplication (quadratic pair representations, cubic compute).
This setup isolates whether performance gains stem from the richer pair representation or from cubic-order computation.

Across data regimes, all architectures struggle under extreme data scarcity.
However, as data and depth increase, standard self-attention consistently lags behind both cubic-compute methods (see~\reffig{match3}).
Notably, triangle multiplication on quadratic pair representations outperforms both variants of Transformer-style attention in shallow settings, demonstrating a particular advantage in capturing nonlocal 3-token interactions even when computational budgets are matched.

\lblsec{match3}
\begin{figure}[h]
\centering
\includegraphics[width=0.8\linewidth]{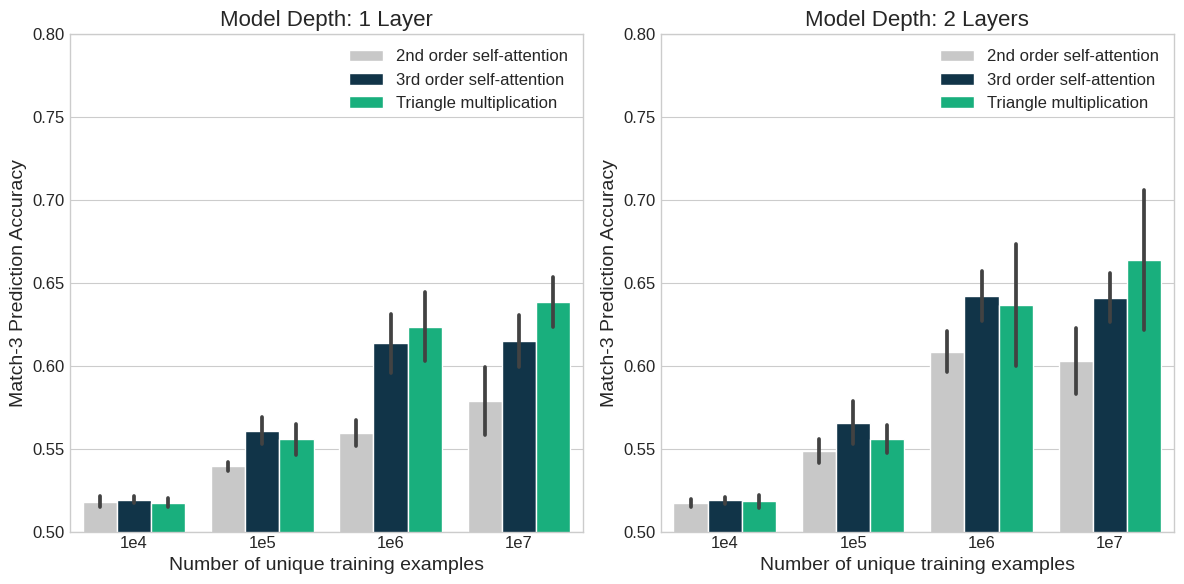}
\caption{
\textbf{Comparison between architecture variants on Match3.}
We report classification accuracy on Match3 task as a function of training data size and model depth.
}
\lblfig{match3}
\vspace{-1pt}
\end{figure}

\clearpage

\subsection{Dissecting Pairmixer performance on the RCSB test set}

To gain deeper insight into these results, we analyze Pairmixer, Pairformer, and Transformer performance across the RCSB test set through two complementary perspectives.

We first examine the correlation between Pairformer and Pairmixer performances. In~\reffig{win_rates_vs_pairformer}, each point represents a single structure, with coordinates indicating the respective model's performance.
The strong correlation between the two models, with minimal outliers, suggests the two architectures share similar failure and success modes.
Furthermore, Pairmixer outperforms Pairformer on lDDT in 44.9\% of the test cases, demonstrating near-equivalent predictive capability despite the architectural simplification.

Next, we investigate whether Pairmixer's competitive performance is confined to favorable conditions, specifically, short sequences or proteins with abundant homologous sequences.
We partition the RCSB test set by sequence length and MSA depth, then evaluate all three models across these stratified subsets in~\reffig{stratified}.
As expected, all models achieve higher accuracy on shorter proteins and those with richer MSA information, while accuracy degrades for longer sequences and sparser alignments.
Critically, these performance trends remain consistent across architectures, with Pairmixer maintaining parity with Pairformer across all difficulty regimes.

\begin{figure}[h]
\centering
\includegraphics[width=0.99\linewidth]{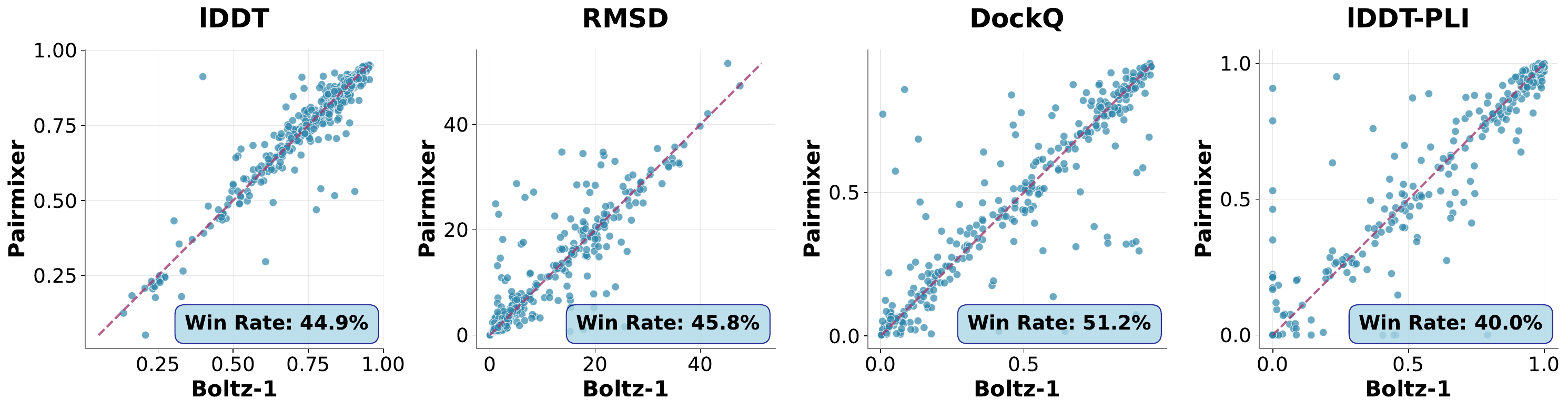}
\caption{
\textbf{Head-to-head comparison between \name and the Pairformer backbone.}
The win rate shows how often the \name architecture achieves a better score than the Transformer architecture.
In 89\% of cases, the two models’ lDDT scores differ by less than 5 points.
}
\lblfig{win_rates_vs_pairformer}
\end{figure}

\begin{figure}[h]
\centering
\vspace{0pt}
\begin{subfigure}[b]{0.99\linewidth}
    \centering
    \includegraphics[width=\linewidth]{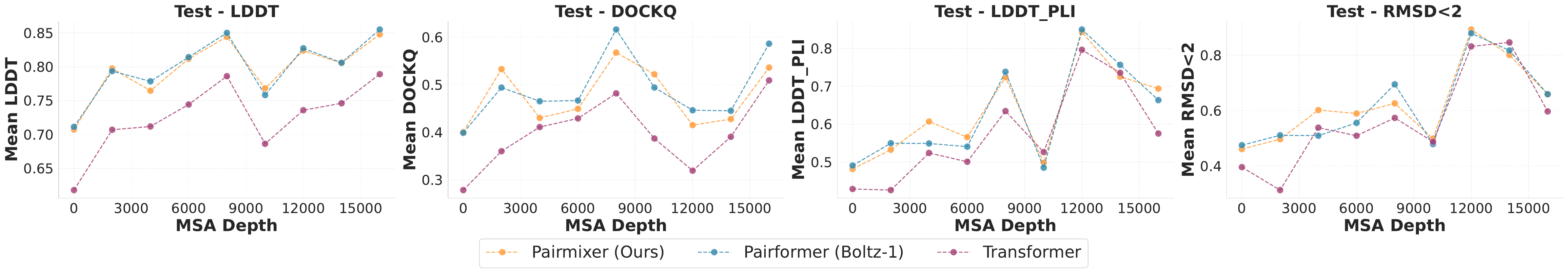}
    \caption{Performance stratified by MSA depth.}
    \lblfig{msa_depth}
\end{subfigure}
\vspace{5pt}
\begin{subfigure}[b]{0.99\linewidth}
    \centering
    \includegraphics[width=\linewidth]{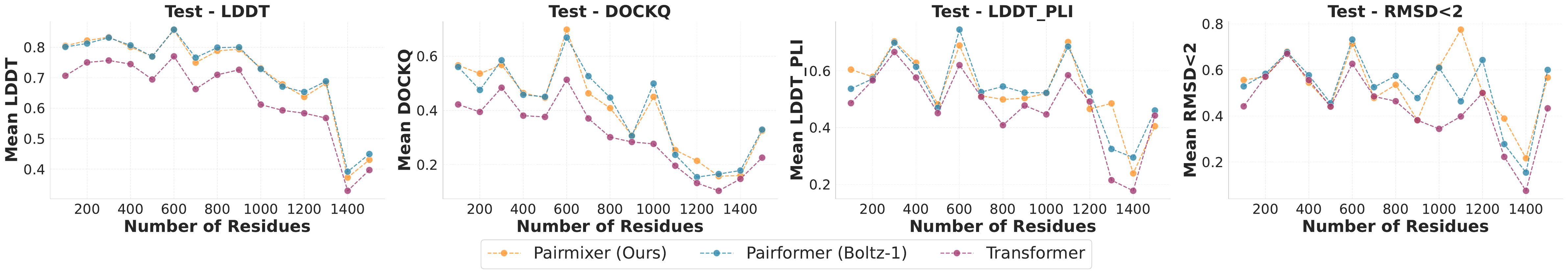}
    \caption{Performance stratified by sequence length.}
    \lblfig{residue_count}
\end{subfigure}
\caption{
\textbf{Performance across different data difficulty metrics.} 
Pairmixer maintains comparable performance to Pairformer across all difficulty levels.
}
\lblfig{stratified}
\vspace{-1pt}
\end{figure}

\clearpage

\subsection{Distogram Performance}
A potential confounding factor in evaluating Pairmixer is that the diffusion module may correct errors in lower-quality backbone outputs, producing high-quality structures independently of the backbone.
To isolate the backbone’s contribution, we evaluate distogram predictions on the RCSB test set.
The distogram head predicts a 64-bin discretized distance matrix directly from backbone features $\vz^{\text{backbone}}$, and accuracy is measured both across the full system and between chains in~\reftbl{distogram_accuracy}.
Following standard contact prediction evaluation~\citep{moult2014casp}, we also report precision at $L$ and $L/5$ in~\reftbl{distogram_precision}.
Across all metrics, \name performs comparably to Pairformer, suggesting that the Pairmixer with triangle multiplication and feed-forward networks produce equally expressive backbone features.

\begin{table}[htbp]
\centering
\caption{\textbf{Distogram Prediction Performance.}}
\lbltbl{distograms}
\begin{subtable}[t]{\linewidth}
\centering
\caption{Global and Inter-Chain Accuracy}
\lbltbl{distogram_accuracy}
\begin{tabular}{lllll}
Method & \multicolumn{2}{c}{Global} & \multicolumn{2}{c}{Inter-Chain} \\
& Top-1 Acc & Top-5 Acc & Top-1 Acc & Top-5 Acc \\
\shline
Pairformer (Boltz-1) & 0.74 & 0.89 & 0.67 & 0.73 \\
Pairmixer (Ours)     & 0.73 & 0.88 & 0.67 & 0.73 \\
Transformer          & 0.72 & 0.86 & 0.67 & 0.72 \\
\end{tabular}
\end{subtable}
\begin{subtable}[t]{\linewidth}
\vspace{11pt} %
\centering
\caption{Contact Prediction}
\lbltbl{distogram_precision}
\begin{tabular}{lcccccc}
Method & \multicolumn{2}{c}{Short} & \multicolumn{2}{c}{Medium} & \multicolumn{2}{c}{Long} \\
 & P@L & P@L/5 & P@L & P@L/5 & P@L & P@L/5 \\
\shline
Pairformer (Boltz-1) & 0.72 & 0.75 & 0.72 & 0.76 & 0.73 & 0.81 \\
Pairmixer (Ours)     & 0.72 & 0.75 & 0.72 & 0.76 & 0.73 & 0.80 \\
Transformer          & 0.69 & 0.72 & 0.69 & 0.74 & 0.70 & 0.79 \\
\end{tabular}
\end{subtable}
\end{table}

\subsection{Sensitivity Analysis to Number of Recycling Steps}
To ensure Pairmixer’s performance is not solely due to greater benefits from recycling, we evaluate Pairformer, Pairmixer, and Transformer with 0, 1, 3, and 10 recycling steps in~\reftbl{recycles}.
All models use the large setting (48 layers) and are trained for the same number of iterations.
Pairformer and Pairmixer achieve similar results across different numbers of recycles, suggesting that Pairmixer’s comparable performance is not simply a result of additional recycling.

\begin{table}[h]
\centering
\caption{
\textbf{Impact of Recycling Steps}.
Performance increases but quickly saturates for all architectures.
}
\begin{tabular}{lcccccc}
Architecture & Recycles & lDDT & $\text{DOCKQ}_{>0.23}$ & $\text{lDDT}_\text{PLI}$ & $\text{RMSD}_{<2}$ \\
\shline
Pairformer (Boltz-1) & 0 & 0.75 & 0.59 & 0.55 & 0.53 \\
Pairformer (Boltz-1) & 1 & 0.77 & 0.59 & 0.58 & 0.56 \\
Pairformer (Boltz-1) & 3 & 0.78 & 0.62 & 0.57 & 0.55 \\
Pairformer (Boltz-1) & 10 & 0.78 & 0.64 & 0.57 & 0.54 \\
\midrule
Pairmixer (Ours) & 0 & 0.74 & 0.59 & 0.54 & 0.52 \\
Pairmixer (Ours) & 1 & 0.76 & 0.61 & 0.56 & 0.56 \\
Pairmixer (Ours) & 3 & 0.77 & 0.61 & 0.56 & 0.54 \\
Pairmixer (Ours) & 10 & 0.78 & 0.63 & 0.57 & 0.55 \\
\midrule
Transformer & 0 & 0.62 & 0.40 & 0.48 & 0.47 \\
Transformer & 1 & 0.67 & 0.49 & 0.51 & 0.48 \\
Transformer & 3 & 0.70 & 0.52 & 0.52 & 0.49 \\
Transformer & 10 & 0.70 & 0.53 & 0.51 & 0.48 \\
\end{tabular}
\lbltbl{recycles}
\end{table}

\clearpage 

\section{Model Hyperparameters}
\reftbl{boltz1_hyperparameters} includes a thorough list of the hyperparameters used for our experiments.
This table additionally includes the training FLOPs for all model architectures and sizes.

\begin{table*}[h]
\centering
\caption{
\textbf{Model Hyperparameters.}
Dashes (-) indicate same value as the previous column.
The large model is trained with smaller crops and mixed data, then with larger crops and PDB-only data.
}
\vspace{-1em}
\setlength{\tabcolsep}{0.3em}
\renewcommand{\arraystretch}{1.2}
\small
\begin{tabular}{@{}lcccc@{}}
\toprule
\textbf{Hyperparameter} & Small & Medium & Large Stage 1 & Large Stage 2 \\
\midrule
\multicolumn{5}{l}{\textit{\textbf{Model Architecture}}} \\
\midrule
Number of Backbone Layers       & 12  & 24 & 48 & 48 \\
Number of MSA Layers            & 4   & -  & -  & -  \\
Token representation dim ($C_s$) & 384 & -  & -  & -  \\
Pair representation dim ($C_z$) & 128 & -  & -  & -  \\
Backbone dropout                & 0.25 & -  & -  & -  \\
MSA Module dropout              & 0.15 & -  & -  & -  \\
Number of Diffusion Layers      & 6   & 24 & 24 & 24 \\
Atom representation dim         & 128 & -  & -  & -  \\
Atom pair representation dim    & 16  & -  & -  & -  \\
\midrule
\multicolumn{5}{l}{\textit{\textbf{Training}}} \\
\midrule
Optimizer              & Adam & - & - & - \\
Maximum learning rate  & $1.8 \times 10^{-3}$ & - & - & - \\
Diffusion multiplicity & 16 & - & - & - \\
Recycling              & 0,1,2,3 & - & - & - \\
Epochs                 & 68  & 68 & 68 & 20 \\
Training Samples       & 6.8M & 6.8M & 6.8M & 2M \\
\midrule
\multicolumn{5}{l}{\textit{\textbf{Data Processing}}} \\
\midrule
Data source            & \small{PDB + OpenFold} & - & - & PDB \\
Maximum tokens         & 384 & 384 & 384 & 512 \\
Maximum atoms          & 3,456 & 3,456 & 3,456 & 4,608 \\
Maximum MSA sequences  & 2,048 & - & - & - \\
Samples per epoch      & 100,000 & - & - & - \\
Total Batch size       & 128 & - & - & - \\
\midrule
\multicolumn{5}{l}{\textit{\textbf{Inference}}} \\
\midrule
Number of sampling steps & 200 & - & - & - \\
Maximum MSA Sequences    & 4096 & - & - & - \\
Recycling                & 10  & - & - & - \\
Diffusion samples        & 5   & - & - & - \\
\midrule
\multicolumn{5}{l}{\textit{\textbf{Training Infrastructure}}} \\
\midrule
GPU Type & H200 & - & - & - \\
Number of GPUs & 32 & 32 & 32 & 64 \\
\midrule
\multicolumn{5}{l}{\textit{\textbf{Total Training FLOPs}}} \\
\midrule
Boltz-1 (Pairformer)   & 8.306e+19 & 1.467e+20 & 2.707e+20 & 1.572e+20 \\
\name                  & 4.817e+19 & 8.557e+19 & 1.572e+20 & 8.716e+19 \\
Transformer            & 5.784e+18 & 7.888e+18 & 8.941e+18 & 4.205e+18 \\
\midrule
\multicolumn{5}{l}{\textit{\textbf{Inference FLOPs}}} \\
\midrule
Boltz1 (Pairformer) & 9.100e+15 & 1.595e+16 & 2.964e+16 & - \\
\name               & 4.474e+15 & 7.849e+15 & 1.460e+16 & - \\
Transformer         & 4.137e+14 & 4.975e+14 & 6.652e+14 & - \\
\end{tabular}
\lbltbl{boltz1_hyperparameters}
\end{table*}

\end{document}